# Messaging-based Adaptive Vector Computing (MAVeC) Accelerator for AI Workloads

Md Rownak Hossain Chowdhury*

University of Missouri-Kansas City (UMKC), Kansas City, MO, US, rhctmc@umkc.edu

Mostafizur Rahman

University of Missouri-Kansas City (UMKC), Kansas City, MO, US, rahmanmo@umkc.edu

The performance of AI accelerators is increasingly limited by data movement, memory access, and orchestration overheads rather than raw compute capability. This paper presents MAVeC, a messaging-based adaptive vector computing accelerator designed to support streaming execution and runtime configurability for AI workloads. MAVeC replaces centralized control with a message-driven execution model in which data and control propagate together across distributed hardware elements, enabling autonomous execution, flexible routing, and efficient coordination. We validate MAVeC's core hardware constructs and execution using matrix multiplication and convolution workloads under a cycle-accurate, system-level ASIC design in TSMC 28 nm, capturing computation, communication, and reduction. MAVeC sustains ≥97% array utilization across hardware scales and problem sizes by translating spatial capacity into effective computation. Once inputs are brought in, over 90% of communication remains on-chip through coordinated temporal reuse, spatial multicast, and on-fabric partial-sum reduction. On a 64x64 SiteO array, MAVeC sustains over 5 TFLOPs/s while reducing end-to-end latency. Compared to TPU-style systolic arrays and MEISSA under compute-centric models, MAVeC achieves 1.5-2x lower latency. When evaluated against optimized NVIDIA H100 FP32 kernels, MAVeC sustains 5.8–6.1 TFLOPs/s, delivering a consistent 6.0–7.2x throughput advantage across problem sizes. Energy results show that MAVeC converts higher instantaneous power into lower total energy by shortening execution time and amortizing data movement. These results demonstrate that message-driven execution provides an effective architectural foundation for overcoming data-movement and orchestration bottlenecks, enabling scalable, high-utilization accelerators for future AI workloads.

* Place the footnote text for the author (if applicable) here.

## 1 INTRODUCTION

The advent of Artificial Intelligence (AI), particularly deep neural networks (DNNs), has driven a paradigm shift across domains such as computer vision, natural language processing, robotics, and data analytics [2, 21]. As AI models continue to grow in scale and complexity, their execution increasingly depends on parallel, distributed computation and sustained data movement across many processing elements[43, 52]. These characteristics expose a fundamental mismatch with traditional load–store, von Neumann-style computing systems, which rely on centralized memory access and sequential control [19]. Addressing modern AI workloads therefore requires architectural approaches that natively support distributed execution and streaming dataflow[29].

Recent research has proposed a variety of specialized hardware accelerators to address the computational demands of AI workloads, including TPU[23], MEISSA[5], SpArch[51], MatRaptor[39], and CGRA-based inference engines[30]. While these architectures provide substantial performance gains over CPUs and GPUs, many rely on static dataflows, fixed interconnects, or centralized control mechanisms that limit adaptability across diverse and evolving AI models. As workloads increasingly combine dense computation with complex data movement and coordination requirements, the ability to dynamically orchestrate execution becomes a critical architectural concern[18].

In this work, we present messaging-based adaptive vector computing (MAVeC), a programmable accelerator architecture designed around message-driven execution and streaming dataflow. MAVeC organizes computation using distributed compute elements, referred to as SiteOs, which coordinate execution through messages that carry both data and control information. In MAVeC, messages explicitly determine both the next operation and the destination of computation, enabling self-propagating data and instruction streams without centralized scheduling. This messaging-based execution model provides programmability through the structure and flow of messages themselves, allowing dynamic orchestration of computation while reducing dependence on global memory access. The architecture employs a hierarchical and modular organization that enables MAVeC to scale while preserving its execution semantics.

The primary focus of this paper is the validation of MAVeC's core hardware constructs and execution model using representative AI primitives. We demonstrate the operation of the architecture through matrix multiplication and convolution workloads, which form the foundation of many AI models, and analyze key operational characteristics including memory usage, data movement behavior, and the numerical and performance advantages enabled by messaging-based self-propagation. In our ongoing other work[14, 15], we are currently exploring applications, data mapping and memory orchestration. In this paper, we show synthesis and simulation at a TSMC 28 nm technology node and evaluation of MAVeC's area, power, and performance and compare its results against state-of-the-art accelerator architectures. The key contributions of this paper are:

- Details of MAVeC's architecture and core components.
- Messaging based virtual interconnection and demonstration of matrix multiplication and convolutions.
- Analysis of memory usage, data movement, and numerical advantages enabled by messaging-based execution for matrix operations.
- Quantitative evaluation of area, power, and performance at TSMC 28 nm technology node for various foundational workloads, and comparison to existing accelerators.

This paper is organized as follows. Section 2 reviews prior work in reconfigurable hardware architecture and establishes the motivation for specialized AI accelerators. Section 3 delves into the MAVeC architecture with system-level overview and demonstration of hardware hierarchy. Next, Section 4 outlines the methodology for performing matrix multiplication and convolution operations, describing the data mapping strategy within the MAVeC fabric. Section 5 introduces the analytical performance modeling framework, which derives utilization, clock-cycle, latency, throughput, and energy formulations to characterize MAVeC behavior across workloads and hardware configurations. Section 6 describes the implementation and validation of MAVeC in TSMC 28nm technology and evaluates key design metrics, including resource utilization, power, latency, and throughput, along with a comparison against state-of-the-art architectures. Finally, section 7 summarizes the concluding remarks emphasizing key contributions of the MAVeC.

## 2 PRIOR WORK & MOTIVATION

The merit of an AI accelerator hinges on the balance between efficiency and flexibility[47]. Hence, accelerator performance lies in its ability to support diverse AI models while sustaining high throughput and energy efficiency across a wide range of neural network configurations[40]. This trade-off has driven AI accelerator research towards two distinct computing paradigms: (i) Reconfigurable computing[4, 24] and (ii) Task specific Specialized computing. Specialized architecture is highly optimized for dedicated tasks, but they often fall short in addressing the broader spectrum of application requirements. For instance, SpArch[51] and MatRaptor[39] are fine-tuned for highly sparse networks like the Amazon co-purchase network[28], while MEISSA[5] leads in low-latency tasks in edge devices like autonomous drones, and TPUs[23] dominate high-throughput scenarios like data centers. Despite their strengths, rigid, application-centric design approach leads to shorter lifecycles and increased non-recurring engineering costs when deploying different AI workloads[34].

In contrast, reconfigurable architecture provides flexibility by dynamically adapting hardware to varying computational tasks[42]. Several reconfigurable paradigms-including Coarse-Grained Reconfigurable Architectures (CGRAs)[33], Field-Programmable Gate Arrays (FPGAs)[26], memristor-based reconfigurable circuit[49] have been explored to address this challenge. However, this flexibility often comes at the cost of architectural overhead. For example, FPGAs offer fine-grained, gate-level flexibility but suffer from lower clock frequencies and longer configuration times[35]. CGRAs attempt to balance flexibility and efficiency through coarse-grained reconfiguration, positioning themselves between FPGAs and rigid, task-specific ASICs[32]. CGRAs are therefore well-suited for data-parallel workloads, including digital signal processing, image processing, and machine learning[6]; however, their effectiveness diminishes for modern deep learning workloads. The CGRA-based VersatCNN architecture[22], for example, incurs substantial overhead due to frequent reconfiguration between convolution, max-pooling, and up-sampling layers, while its fixed parallelism of 832 MAC units leads to underutilization in layers with lower computational intensity or irregular dataflows[48]. The CGRA-based EMAX architecture proposed by Tanomoto et al.[41] is centered around a grid of PEs with fixed 32-bit, 2-stage floating-point units (EX1, EX2), FIFOs, and local memory to facilitate parallelism. However, the fixed execution units limit efficiency for irregular operations such as non-linear activations and pooling operations, leading to underutilization of hardware resources[31]. Although EMAX supports flexible operation mapping, its largely static configuration lacks dynamic reconfigurability, limiting adaptability to diverse and evolving CNN tasks[3]. Moreover, while pipeline parallelization is applied to both forward and backward propagation, the fixed PEs and the double-buffering mechanism introduce synchronization overhead for larger convolutional kernels, resulting in performance bottlenecks in deeper neural networks. Similarly, the reconfigurable hybrid neural-network processor proposed by Yin et al.[50] achieves energy efficiency of 1.06-to-5.09 TOPS/W but at the expense of operating frequency (as low as 10 MHz) and real-time throughput[46]. The use of heterogeneous processing elements to support specialized operations like pooling and RNNs increases architectural complexity, necessitating dynamic PE partitioning. This external host-controlled partitioning incurs significant overhead, limiting adaptability in real-time environments.

The Eyeriss architecture[10], though optimized for energy efficiency, exhibits limitations in both reconfiguration overhead and data movement. The 1794-bit scan chain for reconfiguring the system between CNN layers introduces overhead, limiting real-time adaptability in deep networks that require frequent reconfiguration[11]. More broadly, accelerator architectures that rely on Network-on-Chip (NoC) communication to reduce off-chip data movement often suffer from synchronization delays during large convolutional layers, such as those in VGG-16, leading to reduced throughput[27]. Collectively, prior accelerator designs reveal a fundamental trade-off between efficiency, flexibility, and data-movement efficiency. Specialized architecture achieves high performance for fixed workloads but lack flexibility, while reconfigurable and CGRA-based systems suffer from compiler-driven control overhead, frequent reconfiguration, synchronization delays, and inefficient data movement. As AI workloads grow increasingly diverse and memory-dominated, these limitations hinder sustained utilization and throughput. These limitations motivate a runtime-reconfigurable execution model that sustains high utilization while minimizing control and data-movement overhead - forming the foundation of the proposed Messaging-based Adaptive Vector Computing (MAVeC) architecture.

## 3 MAVEC ARCHITECTURE

Messaging-based Adaptive Vector Computing (MAVeC) is a reconfigurable accelerator built on a deterministic, message-driven execution model that eliminates centralized control. Computation, data operands, and control semantics are encoded directly into messages, which autonomously propagate through the hardware fabric. This source-destination abstraction enables dynamic execution paths over a fixed physical substrate, allowing MAVeC to combine ASIC-like efficiency with programmability. Figure 1(a) illustrates the system-level organization of MAVeC. A host CPU interfaces with the accelerator through PCIe, translating high-level programs into MAVeC messages and collecting results. Internally, MAVeC consists of a hierarchical array of compute tiles connected by distributed buses and embedded memory. Majority of data movement, computation, and control within the accelerator are realized exclusively through message exchanges, eliminating centralized scheduling and global synchronization. The remainder of this section presents the MAVeC architecture, spanning message encoding, instruction set, microarchitecture, communication and routing, and memory organization.

### 3.1 Message Encoding

MAVeC employs a 64-bit message format, shown in Table 1, which serves as the fundamental unit of execution. Each message comprises five fields: Present Opcode (*PO*), Present Address (*PA*), Value, Next Opcode (*NO*), and Next Address (*NA*), and is categorized as one of three classes: execution messages with explicit successor information (*NO*, *NA*), terminal messages that conclude operation chain, or pattern-driven messages used for workload orchestration[15]. Messages can be injected by the host or generated on-chip. In the latter case, MAVeC synthesizes new messages locally as a direct outcome of computation, allowing execution to progress autonomously within the accelerator fabric. When a message reaches its destination, the operation specified by the *PO* is applied to the embedded value, and the resulting value, together with the *NO* and *NA* fields, defines the generated message.[20] Through this chaining mechanism, MAVeC supports instruction sequencing without program counters or centralized control.

Table 1: MAVeC message encoding format

| Bit Position | | 0:3 | 4:15 | 16:47 | 48:51 | 52:63 |
|---|---|---|---|---|---|---|
| Field | Type-1 | PO | PA | Value | NO | NA |
| | Type-2 | PO | PA | Value | 16’b0 | |
| | Type-3 | PO | PA | Value | Workload Pattern | |

### 3.2 Instruction Set Architecture (ISA)

MAVeC defines a compact instruction set consisting of 13 operations, summarized in Table 2, to support message-driven execution across the accelerator fabric. The ISA includes one programming instruction (Prog) for initializing stationary state and twelve execution instructions that implement arithmetic primitives. For each arithmetic operation, MAVeC provides both scalar and streaming variants: scalar instructions store results locally within a SiteO, while streaming instructions forward results to neighboring SiteOs as part of a message stream. Non-linear activation functions are realized through linear approximation using the available arithmetic primitives[37]. Together, these instructions capture the computation and dataflow patterns required by neural network workloads, including convolution, recurrent operations, and activation functions, within the MAVeC execution model.

### 3.3 Microarchitecture

MAVeC follows a scalable hierarchical organization composed of Quads, Blocks, Tiles, SiteMs, and SiteOs. Each Quad contains four Blocks; each Block comprises four Tiles; each Tile integrates sixteen SiteMs; and each SiteM consists of sixteen SiteOs arranged in a two-dimensional grid. This hierarchy enables parallel task decomposition and localized communication, with SiteOs serving as the fundamental processing elements and higher-level structures coordinating message routing and data distribution. Each SiteO is responsible for both computation and message propagation. As shown in Figure 1(c), a SiteO integrates an IEEE-754 floating-point unit (FPU), a message decoder, local registers, two FIFOs (top and left directions), and on-chip SRAM (L0) for weight storage. Messages enter from the top or left and exit toward the right or bottom, with each SiteO storing the addresses of its immediate neighbors to enable fully local routing without global coordination. Incoming messages are checked for destination matching; matching messages are decoded and executed using the FPU, while non-matching messages are forwarded. SiteOs support two execution behaviors: streaming, in which computed results are forwarded as part of a message stream, and forwarding, in which messages are relayed without modification. FIFO-based flow control enables pipelined message handling with a one-cycle pass-through latency when empty and enforces backpressure to prevent buffer overflow. The initial phase during which stationary data are loaded into SiteOs is referred to as the programming phase.

Table 2: MAVeC Instruction Set Architecture

| Instruction | Opcode | Task |
|---|---|---|
| Prog | 0001 | Store weights and routing data |
| UPDATE | 1101 | Update SiteO with incoming data |
| A_ADD | 0100 | Update SiteO after addition |
| A_ADDS | 0111 | Stream addition result to target SiteO |
| A_SUB | 0101 | Update SiteO after subtraction |
| A_SUBS | 1000 | Stream subtraction result target SiteO |
| A_MUL | 0010 | Update SiteO after multiplication |
| A_MULS | 1001 | Stream multiplication result target SiteO |
| A_DIV | 0110 | Update SiteO after division |
| A_DIVS | 1010 | Stream division result target SiteO |
| Av_ADD | 1011 | Update SiteO after averaging |
| RELU | 0011 | ReLU activation operation |
| CMP | 1100 | Update SiteO after comparison |

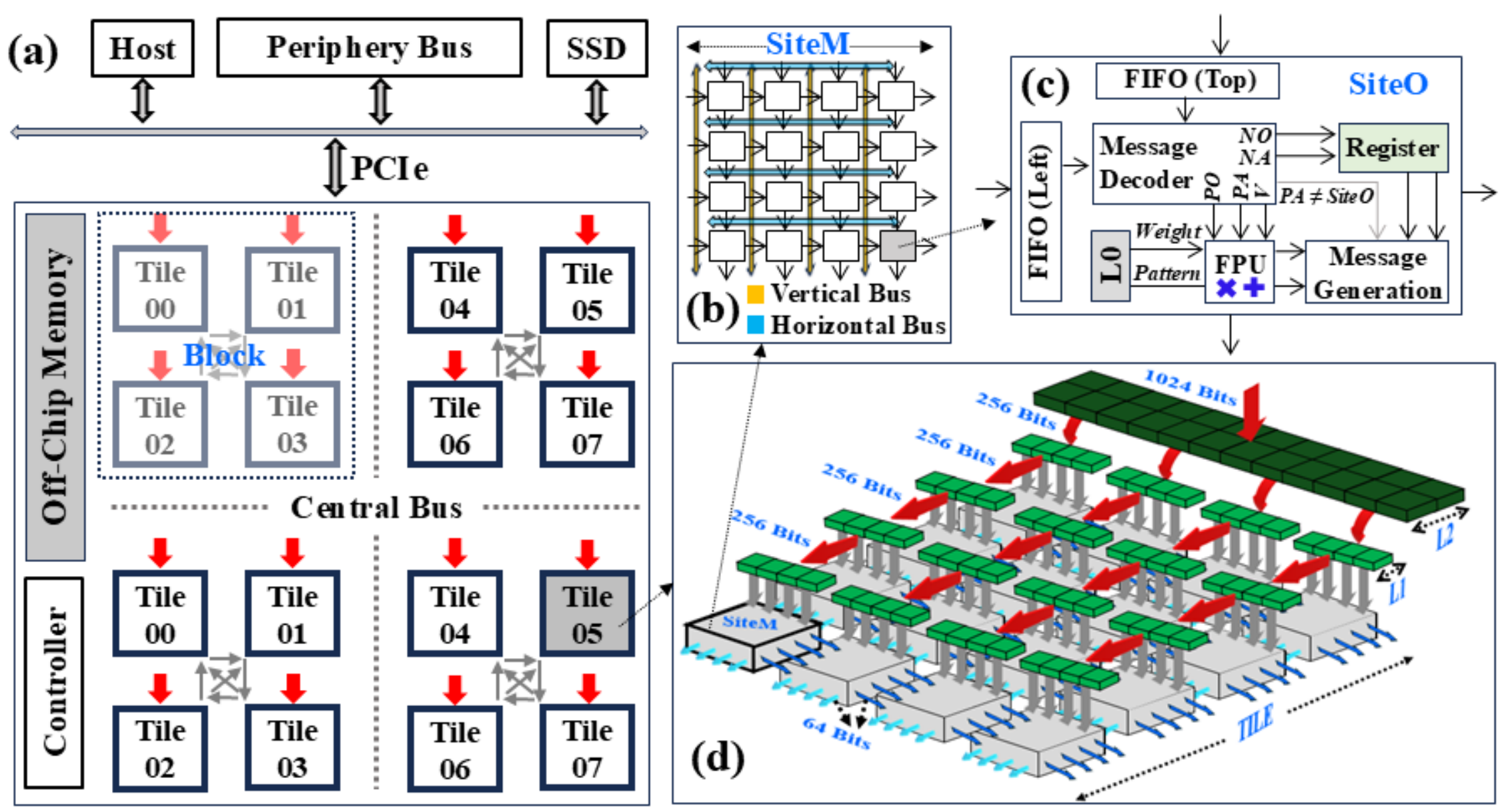

Figure 1: **Overview of the MAVeC architecture. (a)** System-level organization with host, off-chip memory, and accelerator interfacing. **(b)** SiteM organization with SiteO array and horizontal/vertical buses. **(c)** SiteO microarchitecture with FIFOs, message decoding, FPU, and local storage. **(d)** Hierarchical on-chip memory organization across Tiles and SiteMs.

### 3.4 Communication and Routing

Messages are coordinated across SiteOs within a SiteM using horizontal and vertical buses, enabling low-latency communication and simultaneous delivery. Each SiteM is equipped with a vertical bus per column and a horizontal bus per row, as shown in Figure 1(b). The vertical buses provide multicast across SiteO columns, while the horizontal buses support reduction and accumulation across SiteO rows. During the programming phase, messages propagate across SiteM rows through the interconnect to initialize stationary state and establish execution context before computation. Routing decisions are selected based on workload characteristics to maximize data reuse and improve performance[14].

### 3.5 Memory Organization

MAVeC employs a hierarchical memory organization consisting of off-chip memory and a three-level on-chip memory hierarchy, as shown in Figure 1(d). Off-chip memory stores host-generated messages, while on-chip memory is distributed across the accelerator hierarchy to enable localized access and scalable data movement. At the top of the on-chip hierarchy, each Tile integrates a dedicated Tile Memory (L2) of 0.125 KB, yielding 2 KB per Quad across 16 Tiles. Each SiteM is equipped with a private SiteM Memory (L1) of 96 KB; with 256 SiteMs per Quad, this level contributes 24,576 KB. At the lowest level, each SiteO incorporates 64 B of embedded memory (L0) for storing weights and biases, resulting in 64 KB across 1024 SiteOs per Quad. In total, the on-chip memory capacity per Quad is 24,642 KB, corresponding to approximately 98.5 MB across four Quads. Data movement across memory levels is supported by hierarchical buses, including sixteen 1024-bit buses connecting off-chip memory to Tile Memories, sixty-four 256-bit buses between Tile and SiteM Memories, and one hundred and ninety-two 256-bit buses enabling message transfer across SiteM layers[13].

### 3.6 MAVeC Architectural Capability

Table 3 provides a structured overview of the MAVeC architecture, organizing its key design components across system-level integration, computational resources, execution semantics, memory hierarchy, routing fabric, and data orchestration. At the system level, MAVeC is implemented as a PCIe-attached accelerator in TSMC 28 nm, supporting GEMM, convolution, and activation workloads through a message-driven, decentralized runtime model. The hardware fabric follows a hierarchical organization from Quads down to SiteOs, enabling scalable parallel execution with localized control and communication. The execution model is defined through a compact 64-bit message format, a lightweight instruction set, and distinct operational phases that separate weight loading, programming, and computation. MAVeC employs a multi-level and distributed on-chip memory hierarchy with dedicated routing resources at the SiteO, SiteM, and Tile levels to support efficient data movement. Data orchestration mechanisms such as weight-stationary mapping, temporal and spatial reuse, multicast delivery, and sequential hopping are captured at a high level here, while their quantitative impact on utilization, performance, and energy efficiency is analyzed in later sections.

### 3.7 Comparison of MAVeC with other Similar Accelerator Architectures

Domain-specific AI accelerators, such as Google's Tensor Processing Unit (TPU), achieve high efficiency by optimizing fixed dataflows for dense multiply–accumulate operations using systolic arrays[1, 23]. While effective for regular matrix computations, this rigid execution model limits adaptability to diverse tensor shapes, control patterns, and evolving workloads. In contrast, MAVeC adopts a dynamically reconfigurable execution model in which computation, control, and data movement are encoded directly within messages. Each message specifies both the current and next operations along with spatial addresses and data, enabling hardware to execute tasks while autonomously configuring subsequent execution without centralized scheduling.

Compared to Coarse-Grained Reconfigurable Architectures (CGRAs), MAVeC departs from the traditional compiler-driven control paradigm. Conventional CGRAs rely on compiler-orchestrated arrays of processing elements (PEs) with explicit n-to-n routing across reconfigurable interconnects[16, 25, 36, 38]. As system scales, managing register state, routing configuration, and control flow become increasingly complex, particularly for workloads containing loops and conditional branches, often resulting in execution uncertainty and underutilization[8, 9, 45]. MAVeC instead exploits the largely deterministic parallelism of AI workloads by mapping computation as tensor operations and embedding reconfigurability within a message-driven execution model, allowing hardware components to execute autonomously and sequence tasks through on-chip message generation and distributed memory management.

At the microarchitectural level, MAVeC replaces global routing and NoC-based communication with localized, runtime message passing across adjacent tiles using fine-grained hierarchical buses. This approach reduces routing complexity and synchronization overhead compared to NoC-based designs, which depend on predefined paths and routing algorithms for data optimization[7]. By minimizing interconnection overhead, MAVeC enables denser integration of computational units and higher effective parallelism. Furthermore, unlike CGRAs that separate instruction and data memory under a load–store paradigm—often creating fine-grained memory bottlenecks at the PE level—MAVeC unifies data movement and instruction handling within a single messaging-based system, significantly reducing off-chip traffic and improving scalability.

Table 3: Architectural overview of reconfigurable MAVeC accelerator

| SECTION | MODULE | DESCRIPTION |
| --- | --- | --- |
| *System Overview* | Implementation | ASIC (TSMC 28nm) |
| | Reconfigurability | Dynamic adaption via message routing |
| | Runtime Behavior | Message-driven, decentralized control |
| | Workload | GEMV, GEMM, Convolution, Activation |
| *System On Chip (SoC)* | System Integration | PCIe-based host ↔ MAVeC |
| | Peripheral Integration | GPIO, PLIC, UART access via periphery bus |
| *Computational Resources* | Hierarchy | Quad → Block → Tile → SiteM → SiteO |
| | Microarchitecture | SiteO (FPU, decoder, 2 FIFOs, 64 B buffer) |
| *Execution Model* | Message Encoding | 64-bit (PO, PA, OP, NO, NA) |
| | ISA | 13 (1 for Program; 12 for Operation) |
| | Data Types | 1) Weights (32-bits) 2) Messages (64-bits) |
| | Operational Phases | Weights Loading, Programming, Operation |
| | Message Types | 1) Program 2) Operation |
| *On Chip Memory* | Memory Hierarchy | L2 (2 KB), L1 (24 MB), L0 (64 KB) per Quad |
| | Memory Interconnect | 64 (Tile → SiteM); 192 (Intra-SiteM) |
| *On Chip Routing* | SiteO-Level Routing | Top/Left (In) → Bottom/Right (Out) |
| | SiteM-Level Routing | Vertical and Horizontal buses |
| | Tile-Level Routing | Horizontal buses (Tile → Tile) |
| *Data Orchestration* | Dataflow Strategy | Weight Stationary |
| | Data Reuse | Temporal (Weights); Spatial (Inputs) |
| | Data Reduction | Temporal and Spatial (Partial Sums) |
| | Multicast | 1 pass per SiteM level (Input Activations) |
| | Hopping | Sequential (Weights, Partial Sums) |

## 4 MAVEC COMPUTATION

Matrix–matrix multiplication (GEMM) is the dominant computational primitive in modern neural networks and underlies convolution, recurrent layers, and attention mechanisms[12, 44]. MAVeC executes GEMM operation using message-driven execution model mapped onto spatial SiteO arrays. MAVeC execution model retargets this reconfigurable fabric to both GEMM and convolution by altering only the injected message streams and data mappings without any architectural modification. This section details how GEMM is formulated, transformed, and scheduled on MAVeC, describing the data transformation process, fold generation, MatMul block construction, and pipelined on-fabric execution. To ground this execution model in a concrete workload, we then consider a toy neural network that demonstrates how convolution is expressed as GEMM and executed through the same message-driven pipeline.

### 4.1 Matrix Transformation and Folding

Matrix A ($\in \mathbb{R}^{N\ X\ M}$) is first transformed to align with the SiteO array geometry. MAVeC employs interval-based padding along the column dimension ($M$) to regularize data placement and routing. Given an interval parameter $I$, the column dimension is transformed to $M' = \lceil M/I \rceil\ X\ (I+1)$, by inserting one padding column after every $I$ data column, as illustrated in Figure 2(a). The transformed matrix A' $\left(\in \mathbb{R}^{N\ X\ M'}\right)$ is then partitioned into Matrix-A folds based on the available SiteO array dimensions ($R_P\ X\ C_P$). Each Matrix-A fold ($Fold_i^A$) accommodates a fraction of both $N$ and $M'$ dimensions and fits entirely within a single SiteO array instance. The total number of Matrix-A folds ($Total_{A-Folds}$) is given by eq. (1). Matrix B ($\in \mathbb{R}^{M\ X\ P}$) is transformed to align with the folding of Matrix A and the SiteO execution model. To enable column-wise streaming and reuse across stationary A-folds, Matrix B is first transposed, as shown in Figure 2(b). The same interval-based padding is then applied along the transformed column dimension, yielding B' $\left(\in \mathbb{R}^{P\ X\ M'}\right)$. The transformed matrix B' is partitioned into Matrix-B blocks, each corresponding to one Matrix-A fold; hence, the total number of Matrix-B ($Total_{B-Blocks}$) blocks equal the number of Matrix-A folds, as shown in eq. (2). Within a block, B' is further decomposed into P Matrix-B folds, where each fold ($Fold_j^B$) represents a single output column and is streamed sequentially during execution.

$$\mathrm{Total}_{\mathrm{A-Folds}} = \left\lceil \frac{\mathrm{N}}{\mathrm{R_P}} \right\rceil * \left\lceil \frac{\mathrm{M'}}{\mathrm{C_P}} \right\rceil \tag{1}$$

$$\mathrm{Total}_{\mathrm{B-Blocks}} = \mathrm{Total}_{\mathrm{A-Folds}} \tag{2}$$

ALGORITHM 1: Tiling for MAVeC GEMM Execution

INPUT: Input Matrix {A (N X M), B (M X P)}; Interval I; SiteO Array (RP X CP)
OUTPUT: Set of MalMul Blocks {MMi}
1: $\mathbf{A'}(\mathbf{N\ X\ M'})$ //Transform Matrix A
2: $\mathbf{A_{Fold}} = \{\mathbf{A_0},\ \mathbf{A_1},\ \dots\dots,\ \mathbf{A_i}\}$ //Partition $\mathbf{A'}$
3: $\mathbf{B'}(\mathbf{P\ X\ M'})$ //Transform Matrix B
4: $\mathbf{B_{Block}} = \{\mathbf{B_0},\ \mathbf{B_1},\ \dots\dots,\ \mathbf{B_i}\}$ //Partition $\mathbf{B'}$
5: for $i$= 0 to $|\mathbf{A_{Fold}}| - 1$ do: // $|\mathbf{A_{Fold}}| = |\mathbf{B_{Block}}| = |\mathbf{MM_{Block}}|$
6: $\mathbf{B_i} = \{\mathbf{BF_0},\ \mathbf{BF_1},\ \dots\dots,\ \mathbf{BF_{P-1}}\}$ //Generate Matrix B Flods
7: $\mathbf{MM_i} = \{\mathbf{A}_i,\ \mathbf{BF_{i,0}},\ \mathbf{BF_{i,1}},\ \dots\dots,\ \mathbf{BF_{i,P-1}}\}$ //MatMul Block
8: $\mathbf{PS_i} \leftarrow$ Execute ($\mathbf{MM_i}$) //Execute MatMul Block
9: end

### 4.2 MatMul Block Formation

MatMul blocks constitute the fundamental execution units for GEMM on MAVeC. A MatMul block binds one stationary Matrix-A fold with a corresponding Matrix-B block and produces a partial sum fold. As formalized in Algorithm 1 (Steps 2-4), each Matrix-A fold $A_i$ is paired with a Matrix-B block $B_i$, establishing a one-to-one correspondence. This organization enables reuse of each stationary A-fold across all output columns associated with the same folded $M'$ dimension. Within each Matrix-B block, the matrix is decomposed into $P$ Matrix-B folds, each representing a single output column (Algorithm 1, Step 6). Accordingly, a MatMul block $MM_i$ is defined as $\{A_i, BF_{i,0}, BF_{i,1}, \ldots\ldots, BF_{i,P-1}\}$, where $BF_{i,j}$denotes the $j^{th}$ Matrix-B fold associated with $A_i$ (Algorithm 1, Step 7). During execution (Algorithm 1, Step 8), the Matrix-A fold remains stationary within the SiteO array, while the Matrix-B folds are streamed sequentially. Each streamed Matrix-B fold triggers a multiply–accumulate operation across the SiteO array, producing a partial-sum fold column that corresponds to a subset of output rows and a single output column.

### 4.3 Execution Pipeline for MAVeC GEMM

MatMul block execution couples one stationary Matrix-A fold with its corresponding Matrix-B block to generate a partial-sum fold, as illustrated in Figure 2(c). The execution pipeline progresses through five stages: (1) Matrix-A fold programming, (2) Matrix-B fold multicast and multiplication, (3) Intermediate message propagation, (4) Reserved-column accumulation, and (5) Partial-sum offload. Once the pipeline is filled, these stages operate in steady state across successive Matrix-B folds. Execution begins with Matrix-A fold programming, during which the

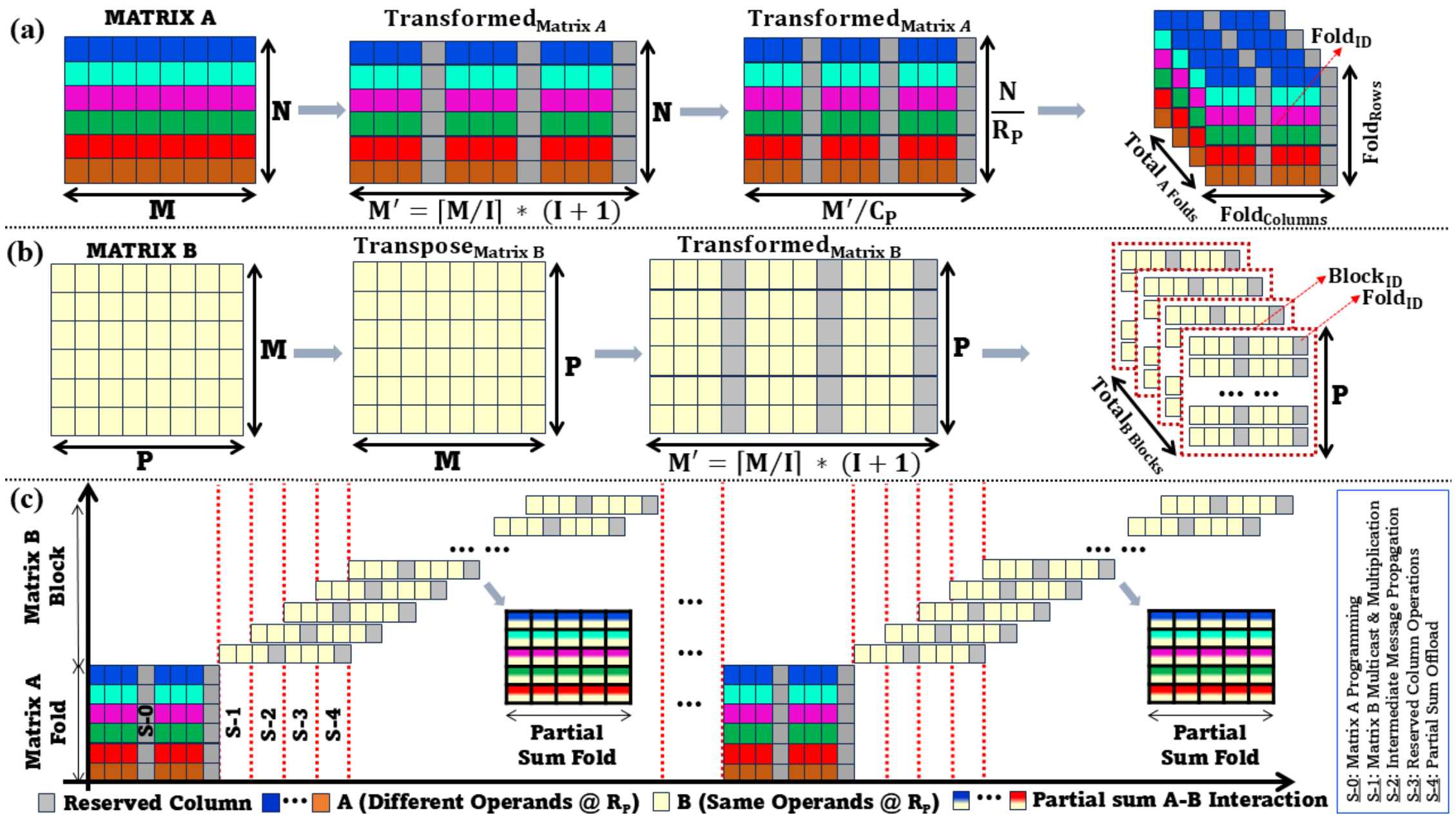

Figure 2: **MAVeC GEMM mapping and execution pipeline. (a)** Interval-based padding and folding of Matrix A into stationary SiteO-sized folds. **(b)** Transposition, padding, and block-wise folding of Matrix B aligned with Matrix-A folds for column-wise streaming. **(c)** Pipelined MatMul block execution showing A-fold programming, B-fold multicast and multiplication, reserved-column accumulation, and partial-sum offload.

A-fold is loaded into the SiteO array and remains resident for the lifetime of the MatMul block. This stationary placement enables the same A-fold to be reused across all $P$output columns associated with the block, amortizing programming overhead and reducing redundant data movement. After programming, the corresponding Matrix-B block—comprising $P$ Matrix-B folds—is streamed sequentially. For each streamed B-fold, operands are multicast across the processing elements and multiplied with stationary A-fold entries. The resulting products accumulate in parallel at the interval-based reserved column locations within the SiteO array. While these partial sums are propagated and reduced through the messaging fabric, subsequent Matrix-B folds are streamed and multiplied concurrently, decoupling multiplication from reduction and offload. Reduced partial sums are finally written to L2 (SiteM) memory. This tightly overlapped pipeline sustains high array utilization, maximizes reuse of stationary data, and minimizes idle cycles, forming the basis for efficient and scalable matrix–matrix multiplication in MAVeC.

### 4.4 Convolution Neural Network

To demonstrate CNN execution on MAVeC, we consider an example network summarized in Table 4 and map it onto a hardware configuration comprising (4X12) SiteO array. SiteO array jointly manages weight programming, data movement, and computation under MAVeC's message-driven execution model. The data mapping on MAVeC follows three principles that align convolution and pooling behavior with its message-driven execution model. First, different filters are programmed across different hardware rows, enabling shared image values to be multicast via the vertical bus and reused across multiple filters, thereby reducing redundant data movement. Second, during weight loading, parameters that traverse longer intra-fabric paths are scheduled earlier to overlap communication latency and avoid synchronization stalls. Image data are grouped according to convolution–pooling dependencies. Each pooling output depends on a localized subset of convolution results, forming multiple independent spatial groups. The input image is therefore partitioned into overlapping groups. Although this introduces redundant computation at group boundaries, parallel execution across SiteO array preserves overall performance.

Figure 3 illustrates the message-driven execution of a convolutional layer on the MAVeC SiteO array. Execution is organized into two phases. In Phase-1 (Weights Loading), convolution filter weights ($F_x^y$) are programmed into SiteOs via hopping, as shown in Figure 3(a). Programming spans four clock cycles, corresponding to the four SiteO rows, after which weights remain stationary. In Phase-2 (Operations), input activations ($I$) are partitioned into four spatial groups dictated by the dependency between a (3X3) convolution and a subsequent (2X2) pooling operation.

Table 4: Network Configuration and performance of an Example Convolutional Neural Network considered for illustration

| NETWORK CONFIGURATION | | | | | | |
|---|---|---|---|---|---|---|
| Layer | | Feature Map | Size | Kernel Size | Stride | Activation |
| IN | Image | 1 | 5X5 | - | - | - |
| 2 | Conv | 4 | 3X3X4 | 3X3 | 0 | RELU |
| 3 | Pool | 4 | 2X2X4 | 2X2 | 1 | RELU |
| 4 | FC | - | 16 | - | - | RELU |
| OUT | FC | - | 4 | - | - | SOFTMAX |

| HARDWARE PERFORMANCE (Precision FP32) | | | | |
|---|---|---|---|---|
| #SiteOs | Freq. (GHz) | Batch | Images/ Batch | Throughput (Images/Second) |
| 48 | 1 | 1 | 20000 | 142.45 X $10^6$ |

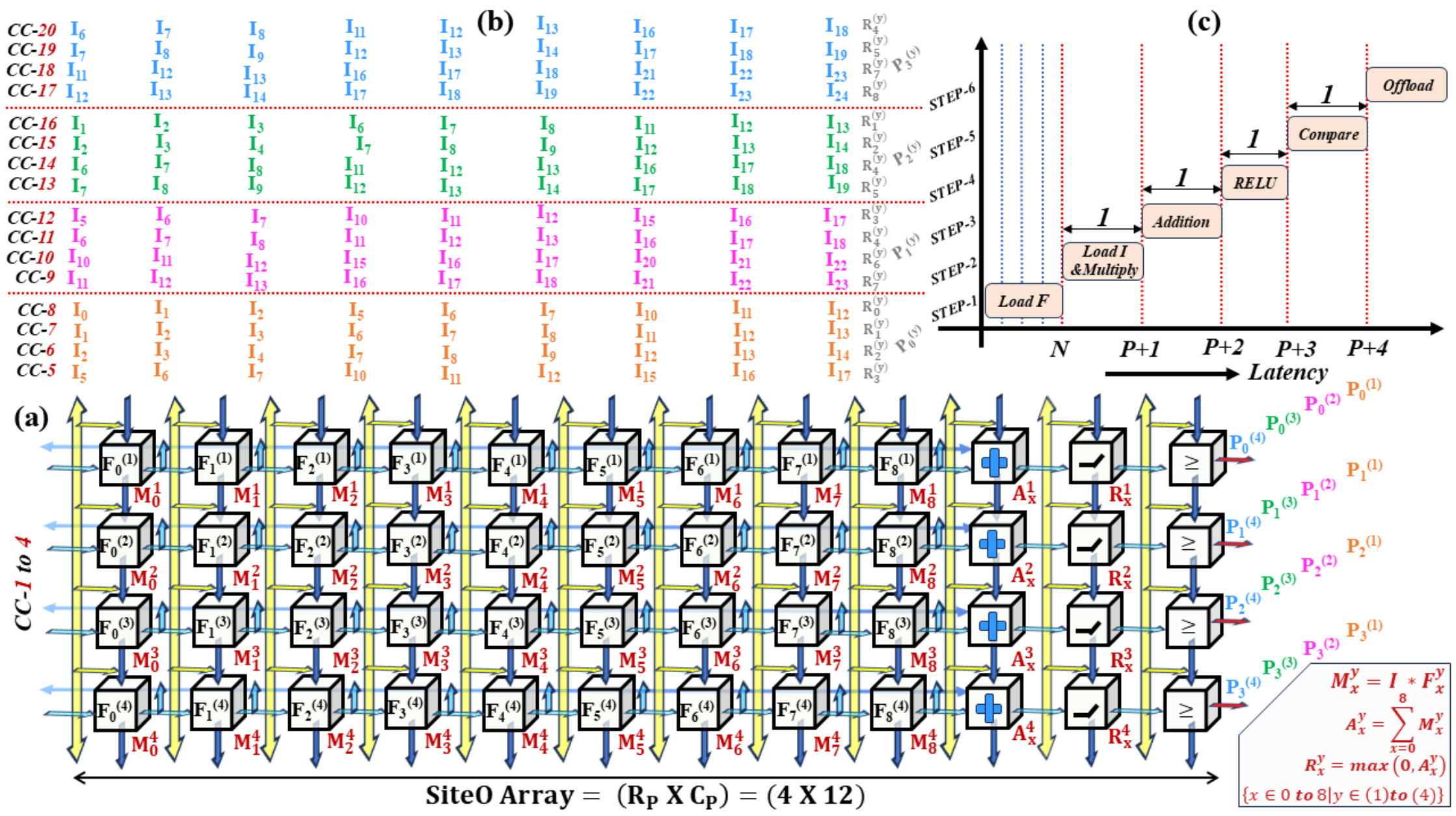

Figure 3: **Message-driven convolution execution on MAVeC. (a)** SiteO array layout (4 X 12) with stationary weights and reserved columns for accumulation, activation, and comparison. **(b)** Pipelined, cycle-by-cycle streaming of grouped input activations across the array. **(c)** Execution timeline showing deterministic progression from weight loading to multiplication, accumulation, activation, comparison, and offload under message-driven control.

Each group contains the activation subset required to generate one pooling output and is streamed into the array in a pipelined manner from CC-5 to CC-20 as shown in Figure 3(b). As activations arrive, parallel multiply operations produce intermediate products ($M_x^y$), which are accumulated to form partial sums ($A_x^y$). These partial sums are then propagated and reduced to generate activation outputs ($R_x^y$), followed by comparison operations that yield pooling outputs ($P_x^y$), as summarized in Figure 3(c). This pipelined flow sustains high utilization, exploits stationary-weight reuse, and enforces a deterministic progression ($\mathrm{M}_\mathrm{x}^\mathrm{y} \rightarrow \mathrm{A}_\mathrm{x}^\mathrm{y} \rightarrow \mathrm{R}_\mathrm{x}^\mathrm{y} \rightarrow \mathrm{P}_\mathrm{x}^\mathrm{y}$) under MAVeC's message-driven execution model.

Figure 4 presents the detailed message schedule for convolution execution on the MAVeC SiteO array. Figure 4(a) (left) lists the incoming programming messages (Phase-1), where Type-1 messages (*PO = Prog*) carry filter values along with next-opcode (*NO*) and next-address (*NA*) information and are routed via hopping to their designated SiteOs in four clock cycles. Figure 4(a) (right) shows the incoming Type-2 operation messages ($PO = A_MULS$), where each message carries image values of different groups and is delivered over successive clock cycles (5-20). The value field encodes the operands to be multiplied, while the next-opcode (*NO*) and next-address (*NA*) fields are set to zero, indicating that these messages rely on locally stored control information (*NO*, *NA*) at the destination SiteOs to enable on-fabric message generation. Each *A_MULS* message is multicast across SiteO rows, enabling parallel multiplication across multiple processing elements within the same cycle window. Figure 4(b) summarizes the spatial location of SiteOs and the logical mapping of intermediate products ($M_x^y$) to partial sums ($A_x^y$), activation outputs ($R_x^y$), and pooling outputs ($P_x^y$). Figure 4(c) illustrates on-chip message generation during Phase-2, where incoming Type-2 operation messages trigger multiplication operations (*A_MULS*) to produce intermediate products. Each result is encapsulated into a new message by attaching the locally stored NO and NA, allowing messages to autonomously propagate through the SiteO array and invoke successive addition, activation, and comparison operations at their destination SiteOs. Through this message chaining, computation progresses deterministically without centralized control or explicit synchronization[14, 15].

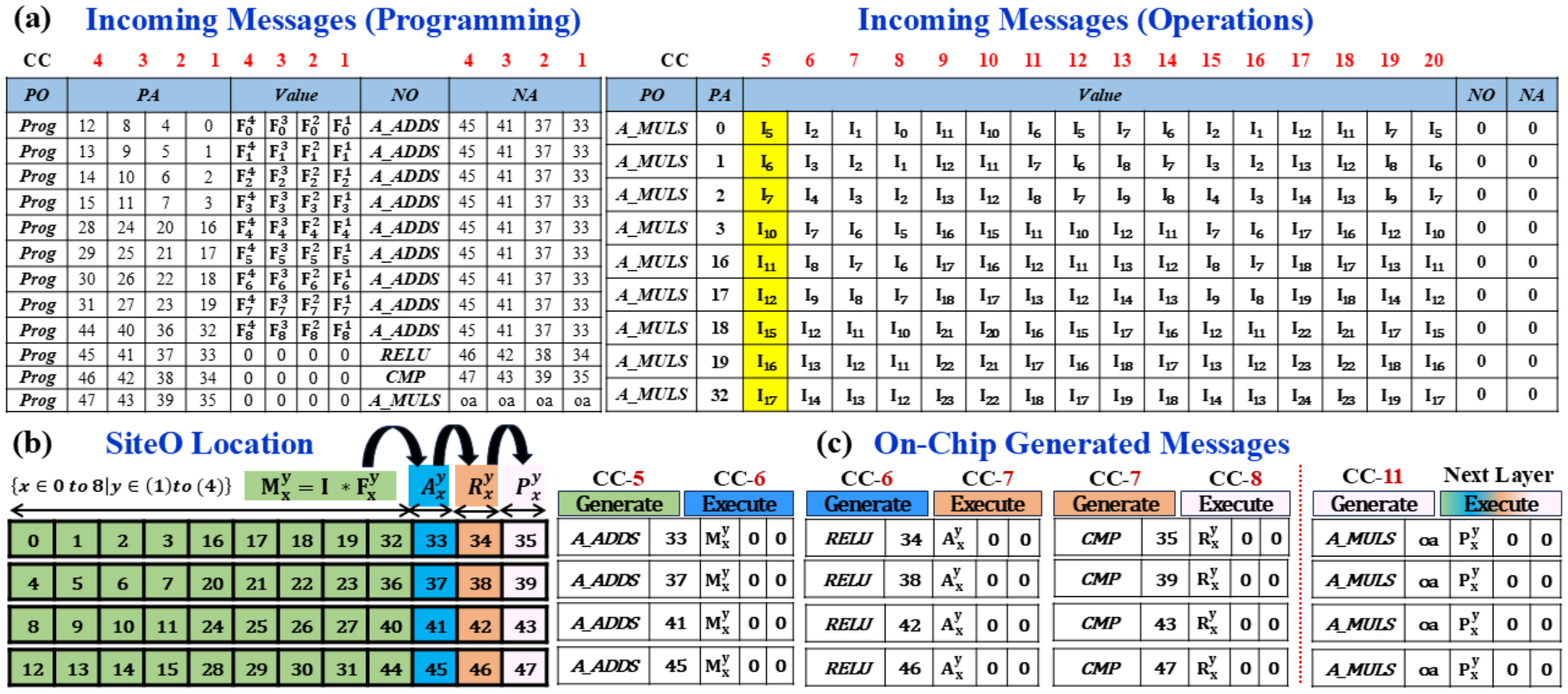

Figure 4: **Message scheduling and on-chip message generation. (a)** Incoming programming and operation messages across clock cycles. **(b)** SiteO spatial layout and logical mapping of intermediate products to partial sums, activations, and pooling outputs. **(c)** On-chip message generation and chaining that autonomously advances computation.

## 5 PERFORMANCE MODELING FRAMEWORK

Section 5 develops an analytical performance modeling framework for MAVeC that quantifies utilization, execution cycles, data orchestration, and energy consumption under its message-driven execution model. The framework is parameterized by array dimensions, tile count, data precision, and workload shape, enabling systematic analysis across both matrix multiplication and convolution without architectural reconfiguration. By decomposing execution into fold-level interactions, the model explicitly accounts for computation, message propagation, and data movement across the hierarchical fabric. Unlike compute-centric models, this framework exposes reconfiguration and orchestration costs intrinsic to MAVeC's execution model, providing architecture-aware insight into how data orchestration decisions translate into measurable performance and energy outcomes. These models form the basis for the evaluations presented in section 6.2.

### 5.1 Average Utilization

Average utilization measures how effectively the SiteO array resources are exercised during matrix multiplication execution. For each MatMul instance $i$, utilization is determined by the number of active processing elements used by the corresponding Matrix-A fold, normalized by the total available array capacity $R_P \ X \ C_P$. Let $Fold_i^A$ denote the number of active elements in the $i^{th}$ Matrix-A fold mapped onto the SiteO array. The number of idle processing elements ($Idle_i$) for that instance follows directly from eq. (3). Utilization for a single MatMul instance is defined as the fraction of non-idle elements relative to the array size. The overall average utilization ($Util_{avg}$) is then obtained by aggregating utilization across all MatMul instances and normalizing by the total number of MatMul executions, as expressed in eq. (4). For example, when a fold size 64X60 ($Fold_i^A$) is processed on a 64X64 ($R_P\ X\ C_P$) MAVeC array, the number of idle SiteOs ($Idle_i$) from eq. (3) is 256 (64*64 – 64*60). Thus, the resulting utilization ($(R_P\ X\ C_P) - Idle_i/(R_P\ X\ C_P)$) for that fold is 0.9375 (64*64 – 256 / 64*64). This computation is performed for each fold, and the results are aggregated and averaged across all MatMul instances ($Total_{MatMul}$) to obtain the overall average utilization in eq. (4). Overall, these formulation captures utilization variations arising from non-uniform fold sizes caused by matrix dimensions, interval-based padding, and array geometry and reflects how effectively MAVeC converts available spatial capacity into useful computation across the execution stream.

$$\mathrm{Idle_i} = (\mathrm{R_P} * \mathrm{C_P}) - \mathrm{Fold_i^A} \tag{3}$$

$$\mathrm{Util_{avg}} = \frac{1}{\mathrm{Total_{MatMul}}} \sum_{\mathrm{i=1}}^{\mathrm{Total_{MatMul}}} \frac{(\mathrm{R_P} * \mathrm{C_P}) - \mathrm{Idle_i}}{(\mathrm{R_P} * \mathrm{C_P})} \tag{4}$$

### 5.2 Message Count

Message count quantifies the communication activity generated by MAVeC's message-driven execution model and serves as a measure for data-movement pressure across off-chip interfaces and the on-chip fabric. Messages are classified into input messages, which originate from off-chip and inject operands into the SiteO array, and intermediate messages, which are generated and consumed entirely on-fabric during computation and reduction. Input messages consist of Matrix-A and Matrix-B messages. Matrix-A input messages ($Input_{Messages\ (A)}$) correspond to the delivery of each Matrix-A fold ($Fold_i^A$) to the SiteO array. Since each fold is loaded once and remains stationary throughout the execution of its associated Matrix-B block, the total number of Matrix-A input messages is given by eq. (5). Matrix-B input messages arise from streaming the Matrix-B operands: for each Matrix-B block, $P$ Matrix-B folds ($\mathrm{Fold_j^B}$) are injected sequentially. The total number of Matrix-B input messages ($Input_{Messages\ (B)}$) therefore follows eq. (6). Together, these expressions capture all off-chip traffic injected into the system for a given workload.

$$\mathrm{Input_{Messages\ (A)}} = \sum_{\mathrm{i=1}}^{\mathrm{Total_{A-Folds}}} \mathrm{Fold_i^A} \tag{5}$$

$$\mathrm{Input_{Messages\ (B)}} = \sum_{\mathrm{i=1}}^{\mathrm{Total_{B-Blocks}}} \sum_{\mathrm{j=1}}^{\mathrm{P}} \mathrm{Fold_j^B} \tag{6}$$

Intermediate messages are generated entirely on-fabric during execution and do not traverse the off-chip interface. Matrix-A–B intermediate messages ($Intermediate_{Messages\ (A-B)}$) correspond to partial products produced by multiply operations within each Matrix-A fold ($Fold(A)_{Rows}^i\ X\ Fold(A)_{Columns}^i$) as it interacts with the $P$ Matrix-B folds; their total count is given by eq. (7). Partial-sum intermediate messages ($Intermediate_{Messages(PS)}$) arise during on-fabric reduction, where intermediate results are accumulated along the reduction dimension; the total number of such messages is captured by eq. (8). Since each Matrix-A fold participates in exactly one Matrix-B block and produces a single partial-sum fold ($PS - Fold_i$), intermediate messages are generated once per Matrix-A fold, and their total count scales with the number of MatMul block interactions. By separating off-chip input messages from on-fabric message activity, the model enables clear attribution of data-movement costs and exposes MAVeC's effectiveness in limiting off-chip communication.

$$\mathrm{Intermediate_{Messages\ (A-B)}} = \sum_{\mathrm{i=1}}^{\mathrm{Total_{A-Folds}}} \mathrm{P} * \mathrm{Fold(A)_{Rows}^i} * (\mathrm{Fold(A)_{Columns}^i} - 1) \tag{7}$$

$$\text{Intermediate}_{\text{Messages(PS)}} = \sum_{i=1}^{\text{Total}_{\text{PS-Folds}}} \text{PS} - \text{Fold}_i \quad (8)$$

### 5.3 Reuse

Reuse characterizes how MAVeC reduces data movement by exploiting temporal locality, spatial multicast, and on-fabric reduction within its message-driven execution model. We quantify reuse in terms of memory savings (in MB) relative to a no-reuse baseline, aggregated across all MatMul blocks. Temporal reuse ($Reuse_{Temporal}$) arises from keeping each Matrix-A fold ($Fold_i^A$) stationary while it is reused across multiple Matrix-B folds within the same MatMul block. For the *ith* Matrix-A fold, the memory footprint ($mem_A(i)$) is given by eq. (9). If the fold were reloaded for each of the $P_i$ Matrix-B folds, the required memory traffic would scale linearly with $P_i$. MAVeC instead loads each Matrix-A fold once per block, yielding a temporal reuse benefit captured by eq. (10). This term accumulates the savings across all Matrix-A folds and reflects MAVeC's ability to amortize operand delivery through keeping matrix-A fold stationarity.

$$\text{mem}_{\text{A}}(\text{i}) = \frac{\text{Fold}_{\text{i}}^{\text{A}} \,.\, (\text{Precision}/8)}{1024^2} \quad (9)$$

$$\text{Reuse}_{\text{Temporal}} = \sum_{i=1}^{\text{Total}_{\text{A-Folds}}} (\text{P}_{\text{i}} - 1).\, \text{mem}_{\text{A}}(\text{i}) \quad (10)$$

Spatial reuse ($Reuse_{Spatial}$) is enabled by multicast distribution of Matrix-B folds across the rows of the SiteO array. The memory footprint ($mem_B(i)$) of the *ith* Matrix-B fold is given by eq. (11). Without spatial reuse, each fold would be replicated independently across $R_P$ rows. MAVeC transmits each matrix-B fold once and fans it out on-fabric, producing the spatial reuse benefit summarized in eq. (12). This formulation captures the reduction in redundant data movement enabled by hardware multicast.

$$\text{mem}_{\text{B}}(\text{i}) = \frac{\text{Fold}_{\text{j}}^{\text{B}} \,.\, (\text{Precision}/8)}{1024^2} \quad (11)$$

$$\text{Reuse}_{\text{Spatial}} = \sum_{i=1}^{\text{Total}_{\text{B-Blocks}}} (\text{R}_{\text{P}} - 1).\, \text{mem}_{\text{B}}(\text{i}) \quad (12)$$

Spatial reduction ($Reduction_{Spatial}$) captures the elimination of redundant partial-sum traffic through on-fabric accumulation. The memory footprint ($mem_{PS}(i)$) of the partial sums associated with the *ith* reduction instance ($PS - Fold_i$) is defined in eq. (13). In the absence of reduction, partial sums would be propagated across all reduction groups formed along the column dimension ($\lceil A_{Col}^{(i)}/I \rceil$). MAVeC performs accumulation within the fabric, reducing the required memory traffic by the factor expressed in eq. (14). This term reflects the benefit of performing accumulation through on-chip message generation, thereby avoiding repeated on-chip movement. Together, these reuse components link MAVeC's execution model to reduced off-chip data movement and energy consumption.

$$\text{mem}_{\text{PS}}(\text{i}) = \frac{\text{PS-Fold}_{\text{i}} \,.\, (\text{Precision}/8)}{1024^2} \quad (13)$$

$$\text{Reduction}_{\text{Spatial}} = \sum_{i=1}^{\text{Total}_{\text{PS-Folds}}} (\left\lceil \frac{\text{A}_{\text{Col}}^{(\text{i})}}{\text{I}} \right\rceil .\, \text{I} - 1) \text{mem}_{\text{PS}}(\text{i}) \quad (14)$$

### 5.4 Clock Cycle

Matrix-A message propagation captures the clock cycles required to deliver a Matrix-A fold from tile memory (L2) to the SiteO array prior to computation. For each tile, the SiteM memory (L1) is organized into four hierarchical levels, and a Matrix-A fold is forwarded sequentially through these levels during propagation. From each L1 memory level, the fold must be delivered to the SiteO array, which is partitioned into four row groups; consequently, each L1 memory level requires four row-wise hops to reach all SiteO rows. This organization results in a total of sixteen propagation passes per tile, corresponding to the traversal of four L1 memory levels and four SiteO row hops per level. Each pass incurs a small, hop-dependent latency that decreases as the hop distance shortens toward upper SiteM levels and nearer SiteO rows of the L2 (Tile memory). Aggregating these passes yields a per-tile propagation cost that scales linearly with the number of tiles ($N_{Tiles}$). Accordingly, the message-propagation latency for a single Matrix-A fold ($T_{Mes(A-Fold)}$) is modeled as eq. (15). Since each Matrix-A fold is injected once per MatMul instance, the total Matrix-A message-propagation latency ($T_{AMP}$) across the execution stream is given by eq. (16).

$$\mathrm{T_{Mes(A-Fold)}} = 1 + (\mathrm{N_{Tiles}}\ \mathrm{X}\ 16) \tag{15}$$

$$\mathrm{T_{AMP}} = \mathrm{Total_{MatMul}}\ \mathrm{X}\ \mathrm{T_{Mes(A-Fold)}} \tag{16}$$

Matrix-B message propagation models the clock cycles required to deliver a Matrix-B block from tile memory (L2) to the SiteO array prior to computation. Like Matrix-A, Matrix-B blocks traverse the hierarchical SiteM memory (L1) levels during propagation. However, once a Matrix-B fold reaches the SiteO array, it is spatially multicast across all SiteO rows in a single cycle. This multicast capability eliminates row-wise hopping within the SiteO array, in contrast to Matrix-A propagation, where explicit row traversal is required. For each tile, the Matrix-B block is forwarded sequentially through the four L1 memory levels. At each level, the transfer from tile memory to SiteM incurs a hop-dependent latency that decreases as the data moves closer to the tile memory (L2). From SiteM to SiteO, the multicast mechanism ensures that all SiteO rows receive the Matrix-B fold simultaneously, resulting in a constant one-cycle transfer regardless of array height. Consequently, each tile contributes four propagation passes—one per SiteM level—yielding a per-block propagation cost that scales linearly with the number of tiles ($N_{Tiles}$). Accordingly, the message-propagation latency ($T_{Mes(B-Block)}$) for a single Matrix-B block is modeled by eq. (17). Since one Matrix-B block is injected per MatMul instance, the total Matrix-B message-propagation latency ($T_{BMP}$) across the execution stream is given by eq. (18).

$$\mathrm{T_{Mes(B-Block)}} = 1 + (\mathrm{N_{Tiles}}\ \mathrm{X}\ 4) \tag{17}$$

$$\mathrm{T_{BMP}} = \mathrm{Total_{MatMul}}\ \mathrm{X}\ \mathrm{T_{Mes(B-Block)}} \tag{18}$$

Weight propagation models the clock cycles required to distribute 32-bits weights from tile memory (L2) into the embedded SiteO memories (L0) prior to computation. Each SiteO contains 64 bytes of embedded storage, while each weight transmission carries 64 bits; consequently, eight repeated transfers are required to fully populate the embedded memory of a single SiteO. Like Matrix-A message propagation, weight delivery proceeds hierarchically through the four SiteM memory (L1) levels associated with each tile, and from each SiteM level the weights are forwarded across the SiteO array, which is partitioned into four row groups. As a result, weight propagation incurs sixteen passes per tile—corresponding to four SiteM levels and four SiteO row hops per level—with each pass repeated eight times to account for the 64-byte SiteO storage fill requirement. Aggregating these effects yields the per-fold weight-propagation ($T_{W(A-Fold)}$) cost in eq. (19), which scales linearly with both the number of tiles ($N_{Tiles}$) and the fixed propagation passes. Since all weights corresponding to Matrix-A fold need to be propagated, the total weight-propagation latency ($T_{WP}$) across the workload is obtained by scaling the per-fold cost by the total number of Matrix-A folds, as expressed in eq. (20).

$$T_{W(A-Fold)} = \{1 + (8 \text{ X } N_{Tiles} \text{ X } 16)\} \quad (19)$$

$$T_{WP} = \text{Total}_{A-Folds} \text{ X } T_{W(A-Fold)} \quad (20)$$

The computation cost models the clock cycles required to execute one complete interaction between a Matrix-A fold and a Matrix-B block on the SiteO array. Each interaction consists of a fixed sequence of pipeline stages: programming and hopping of the Matrix-A fold into the array, multicast of Matrix-B folds, parallel multiply operations across active processing elements, propagation of intermediate messages, reserved-column operations for accumulation control, and partial-sum offload. The aggregate latency of these stages ($T_{MatMul\ Interaction}$) is captured by the per-interaction cost in eq. (21). The constant terms account for pipeline setup (*5 CCs*), message generation (*2CCs*), and offload overheads (*1 CC*), while the linear term $P$ reflects the sequential streaming of $P$ Matrix-B folds for a given Matrix-A fold. The logarithmic term ($log(C_P)/log(I)$) models the multi-stage addition required to reduce partial products across the array columns, where the reduction depth scales with the number of hardware columns $C_p$ and is normalized by the interval-based grouping factor $I$. Since each Matrix-A fold interacts with exactly one Matrix-B block, the total number of Matrix-A–B interactions equal the total number of MatMul instances ($Total_{MatMul}$). The overall computation latency ($T_{Comp}$) is therefore obtained by scaling the per-interaction cost by the total number of interactions, as expressed in eq. (22).

$$T_{MatMul\ Interaction} = \left(5 + P + 2 + \frac{\log(C_P)}{\log(I)} + 1\right) \quad (21)$$

$$T_{Comp} = \text{Total}_{MatMul} \text{ X } T_{MatMul\ Interaction} \quad (22)$$

Partial-sum merging accounts for the clock cycles required to combine intermediate partial sums generated across successive Matrix-A–Matrix-B interactions. For the first interaction, the initial partial sum is loaded into the execution fabric, incurring a fixed setup cost (*4 CCs*). For each subsequent interaction, the merge process consists of loading the incoming partial sum operand (*4 CCs*), performing a single-cycle addition (*1 CCs*), generating the corresponding intermediate message (*1 CCs*), and offloading (*1 CCs*) the updated partial sum to SiteM memory (L1). This sequence introduces a constant per-merge cost that repeats for all remaining interactions. As a result, the total partial-sum merge latency ($T_{PS-Merge}$) grows linearly with the number of Matrix-A–Matrix-B interactions ($Total_{MatMul}$), excluding the initial setup phase. Eq. (23) captures this behavior by modeling the merge latency as a fixed startup cost followed by a repeated per-interaction overhead.

$$T_{PS-Merge} = 4 + \{(\text{Total}_{MatMul} - 1) \text{X } 7\} \quad (23)$$

The total execution time ($T_{Total}(CCs)$) of a workload on MAVeC is obtained by aggregating the clock cycles consumed by all pipeline components, including weight propagation ($T_{WP}$), Matrix-A and Matrix-B message propagation ($T_{AMP}, T_{BMP}$), computation ($T_{Comp}$), and partial-sum merging ($T_{PS-Merge}$), as expressed in eq. (24). This formulation captures the complete end-to-end execution cost across data orchestration, computation, and reduction. The resulting latency is computed by normalizing the total clock cycles by the operating frequency, as shown in eq. (25). Finally, throughput is defined as the ratio of the total floating-point operations performed to the execution latency, as given in eq. (26). Together, these expressions provide a unified performance model that directly links MAVeC's message-driven execution and pipelined dataflow to end-to-end latency and sustained computational throughput.

$$T_{Total}(CCs) = T_{WP} + T_{AMP} + T_{BMP} + T_{Comp} + T_{PS-Merge} \quad (24)$$

$$Latency = T_{Total}(CCs)/Frequency \quad (25)$$

$$Throughput = FLOPs/Latency \quad (26)$$

### 5.5 Energy

Energy consumption is modeled by decomposing both computation and data movement across the memory hierarchy. Compute and memory-access energy values are obtained from post-synthesis characterization using a commercial 28 nm TSMC technology library, as depicted in Table 5 and are applied uniformly across all workloads and configurations. To account for data movement, memory-access energy at each on-chip level is normalized to an energy/byte metric, enabling uniform comparison and aggregation across different (L2, L1, L0) storage levels. In MAVeC, memory access granularity increases hierarchically, with L0 operating at an 8-byte granularity, followed by 32-byte accesses at L1 and 128-byte accesses at L2. Accordingly, the effective read/write energy per byte for a given memory level is defined in eq. (27) as the ratio of the energy consumed per access to the corresponding access granularity.

$$E_{Memory}^{R/W}(PJ/Bytes) = \frac{E_{Memory}\ (PJ/Operations)}{Access\ Granularity\ \ (Bytes)} \quad (27)$$

Table 5: Post-synthesis energy cost of arithmetic operations and memory accesses under TSMC 28nm technology

| Operation | Add | Multiply | L0(R) | L0(W) | L1(R) | L1(W) | L2(R) | L2(W) |
|---|---|---|---|---|---|---|---|---|
| (PJ/Operation) | 1.52 | 2.64 | 3.36 | 3.36 | 12.76 | 11.73 | 10.92 | 9.63 |

Weight propagation energy accounts for the energy consumed in transferring model parameters from off-chip memory through the on-chip memory hierarchy to the compute fabric. In MAVeC, each weight 32 bits and is first read from off-chip memory into the L2 (Tile Memory), then forwarded to L1 (SiteM Memory), and finally delivered to L0 (Embedded SiteO Memory) for consumption by the SiteO array. The energy cost of propagating a single byte of weight data is therefore modeled as the cumulative read–write energy incurred across each transition in this path, as expressed in eq. (28). The total number of weight elements propagated is obtained by summing the sizes of all Matrix-A folds, as shown in eq. (29), and is converted to bytes based on the weight precision given in eq. (30). The total weight-propagation energy ($E_{Weights}$) is then computed by scaling the per-byte propagation cost by the total volume of weight data transferred, as shown in eq. (31).

$$E_{Weight\ (PJ/Byte)} = (E_{Off-Chip}^{R} + E_{L2}^{W}) + (E_{L2}^{R} + E_{L1}^{W}) + (E_{L1}^{R} + E_{L0}^{W}) \quad (28)$$

$$A_{Weight\ (Element)} = \sum_{i=1}^{Total_{A-Folds}} Fold_i^A \quad (29)$$

$$A_{Weight\ (Byte)} = (A_{Weight\ (Element)} * Precision)/8 \quad (30)$$

$$E_{Weights} = A_{Weight\ (Byte)} \ * E_{Weight\ (PJ/Byte)} \quad (31)$$

Message propagation energy captures the cost of transferring messages (opcode and operand) from off-chip memory into the on-chip fabric. In MAVeC, input messages are injected from off-chip memory and propagated through the L2 (Tile Memory) and L1 (SiteM Memory) before reaching the compute fabric, without being stored in L0 (Embedded SiteO Memory). Accordingly, the per-byte message-propagation energy is modeled as the cumulative read and write energy incurred at each level along this path, as defined in eq. (32). Each message has a fixed length of 64 bits, and the total message volume is computed by scaling the number of injected input messages by this message length, as shown in eq. (33) and eq. (34) for Matrix-A and Matrix-B messages, respectively. The total energy consumed by message propagation is then obtained by multiplying the aggregate message volume by the per-byte propagation cost, yielding the energy terms in eq. (35) and eq. (36).

$$\mathrm{E}_{\mathrm{Message\ (PJ/Byte)}} = \mathrm{E}^{\mathrm{R}}_{\mathrm{Off-Chip}} + \mathrm{E}^{\mathrm{W}}_{\mathrm{L2}} + \mathrm{E}^{\mathrm{R}}_{\mathrm{L2}} + \mathrm{E}^{\mathrm{W}}_{\mathrm{L1}} \tag{32}$$

$$\mathrm{A}_{\mathrm{Message\ (Byte)}} = (\mathrm{Input}_{\mathrm{Messages\ (A)}} * \mathrm{Message\ Length})/8 \tag{33}$$

$$\mathrm{B}_{\mathrm{Message\ (Byte)}} = (\mathrm{Input}_{\mathrm{Messages\ (B)}} * \mathrm{Message\ Length})/8 \tag{34}$$

$$\mathrm{E}_{\mathrm{A-Message}} = \mathrm{A}_{\mathrm{Message\ (Byte)}} * \mathrm{E}_{\mathrm{Message\ (PJ/Byte)}} \tag{35}$$

$$\mathrm{E}_{\mathrm{B-Message}} = \mathrm{B}_{\mathrm{Message\ (Byte)}} * \mathrm{E}_{\mathrm{Message\ (PJ/Byte)}} \tag{36}$$

Computation energy accounts for the energy consumed by arithmetic operations performed within the SiteO array during execution. In MAVeC, computation consists of multiplication and addition operations arising from Matrix-A–Matrix-B interactions and on-fabric reduction. The total computation energy ($E_{Computation}$) is therefore modeled as the weighted sum of these two operation types, as expressed in eq. (37), where the number of additions and multiplications is determined by the workload structure and execution schedule. Per-operation energy costs are obtained from post-synthesis characterization and applied uniformly across all experiments. This formulation decouples compute energy from data movement and allows computation cost to be evaluated independently of memory and communication effects.

$$\mathrm{E}_{\mathrm{Computation}} = (\mathrm{N}_{\mathrm{Additions}} * \mathrm{E}_{\mathrm{Addition}}) + (\mathrm{N}_{\mathrm{Multiplications}} * \mathrm{E}_{\mathrm{Multiplication}}) \tag{37}$$

Partial-sum merge energy captures the cost of combining intermediate results produced across successive Matrix-A–Matrix-B interactions. During execution, each interaction generates a partial-sum message that must be propagated and accumulated with previously computed results. The total data volume associated with these intermediate partial sums is quantified in eq. (38), where the number of generated partial-sum messages is scaled by the message length. As shown in eq. (39), propagation energy accounts for reading partial sums from L1 memory and writing the merged results back to L1, reflecting the fact that partial-sum accumulation is performed locally without off-chip traffic. In addition to data movement, each merge operation incurs arithmetic energy due to the required addition, which is incorporated in eq. (40) by scaling the number of intermediate partial-sum messages with the per-addition energy cost. Together, these terms model the combined communication and computation overhead of partial sum merging while explicitly capturing the benefit of performing accumulation close to the compute fabric.

$$\mathrm{PS}_{\mathrm{Message\ (Byte)}} = (\mathrm{Intermediate}_{\mathrm{Messages(PS)}} * \mathrm{Message\ Length})/8 \tag{38}$$

$$\mathrm{E}_{\mathrm{PS\ (Propagation)}} = \mathrm{PS}_{\mathrm{Message\ (Byte)}} * \left(2\mathrm{E}^{\mathrm{R}}_{\mathrm{L1}} + \mathrm{E}^{\mathrm{W}}_{\mathrm{L1}}\right) \tag{39}$$

$$\mathrm{E_{PS}} = \mathrm{E_{PS\,(Propagation)}} + \left(\mathrm{Intermediate_{Messages(PS)}} * \mathrm{E_{Addition}}\right) \tag{40}$$

The total energy consumption of MAVeC is obtained by aggregating all constituent energy components as expressed in eq. (41). This includes the energy associated with weight propagation across the memory hierarchy, Matrix-A and Matrix-B message propagation, arithmetic computation, and partial-sum merging. Each term corresponds to a distinct stage of the message-driven execution pipeline and is computed using the workload-dependent activity counts derived earlier. By summing these components, the model provides a unified view of both data-movement and compute energy, enabling direct attribution of energy cost to specific phases of execution and facilitating fair comparison across different workloads and hardware configurations.

$$\mathrm{E_{Total}} = \mathrm{E_{Weights}} + \mathrm{E_{A-Message}} + \mathrm{E_{B-Message}} + \mathrm{E_{Computation}} + \mathrm{E_{PS}} \tag{41}$$

## 6 EVALUATION AND COMPARISON

This section evaluates MAVeC using a system-level perspective that jointly accounts for computation, data movement, and execution under its message-driven execution model. All MAVeC results are reported in FP32 precision and are derived from end-to-end execution, including weight loading, message propagation, on-fabric computation, and partial-sum reduction. Performance metrics such as utilization, clock cycles, latency, throughput, and energy are computed using the analytical models introduced in section 5 and validated against post-synthesis parameters from a TSMC 28 nm ASIC implementation. Comparisons against state-of-the-art GPU and TPU architectures are conducted using NVIDIA H100 GPU and different TPU implementations. The GPU baseline is derived from vendor-reported optimized kernels and reflects end-to-end execution. TPU results are based on reported optimized implementations and may employ varying precision formats. However, data movement and orchestration costs are typically abstracted or amortized by the execution model considered in TPU

Table 6: Post-synthesis A) power and B) area breakdown of MAVeC hardware components

A) Power (μW)

| Module | Leakage | Internal | Switching | Total |
|---|---|---|---|---|
| FIFO | 9.11 | 53.10 | 10.50 | 72.70 |
| Compare | 0.32 | 2.22 | 1.61 | 4.14 |
| Addition | 2.08 | 91.20 | 70.20 | 163.48 |
| Multiplication | 4.33 | 129.00 | 68.60 | 201.93 |
| ReLU | 0.07 | 0.39 | 0.21 | 0.67 |
| FPU | 7.37 | 108.00 | 69.60 | 184.97 |
| SiteO | 31.10 | 220.00 | 74.80 | 325.90 |
| SiteM | 533.00 | 3830.00 | 1440.00 | 5803.00 |

B) Area ($\mu m^2$)

| FIFO | Sequential | Inverter | Buffer | Logic | Total |
|---|---|---|---|---|---|
| Compare | 1365.59 | 1.26 | 0 | 270.40 | 1637.24 |
| Addition | 0 | 9.58 | 0 | 71.82 | 81.40 |
| Multiplication | 0 | 27.22 | 0 | 532.98 | 560.20 |
| ReLU | 0 | 20.41 | 0 | 1128.08 | 1148.49 |
| FPU | 0 | 0 | 0 | 16.00 | 16.00 |
| SiteO | 67.03 | 48.38 | 0 | 1823.72 | 1939.14 |
| SiteM | 3547.53 | 45.36 | 0 | 2880.49 | 6473.38 |

implementation. Accordingly, these comparisons are intended to emphasize architectural differences in execution paradigms, particularly the impact of MAVeC's message-driven orchestration and hierarchical data movement rather than to assert cycle-level equivalence across heterogeneous platforms.

### 6.1 Hardware Validation

To validate our design, we implemented a complete digital design flow based on CAD tools under TSMC 28nm technology node. Initially, we designed the RTL/behavioral models of MAVeC and iteratively checked for lint compliance, synthesizability, and functional correctness. Early functional simulations were performed using Xilinx Vivado to validate message sequencing and execution behavior. The design was subsequently integrated into a standard ASIC flow using TSMC's 28 nm CLN28HPC+ process (8 metal layers, 0.9 V supply), targeting 1 GHz operating frequency. For synthesis, the design environment was configured by specifying operating conditions, wire-load models, and system interface characteristics. Operating conditions accounted for nominal voltage, temperature, and process assumptions consistent with the CLN28HPC+ PDK. Timing and area constraints were defined through an SDC file and iteratively refined to meet the 1 GHz timing target. Synthesis was performed using Cadence Genus, with optimization guided by timing and area objectives while preserving design rule constraints. Post-synthesis functional verification confirmed behavioral equivalence with the RTL. Table 6 summarizes the resulting area and power breakdown of key MAVeC components under these synthesis conditions.

Functional correctness of MAVeC's message-driven execution was validated through extensive simulation across a range of workloads. For clarity, Figure 5 presents a representative example of a 3X3 matrix-matrix multiplication. In this experiment, nine Type-1 programming messages initialize the SiteOs with Matrix-A values via hopping. Subsequently, columns of Matrix-B are streamed through the vertical bus using Type-2 operation messages. Each incoming activation triggers local multiplication, followed by in-fabric message generation that propagates partial results to designated SiteOs for accumulation. This process repeats until all columns are processed and the results are produced, matching the expected matrix multiplication outputs.

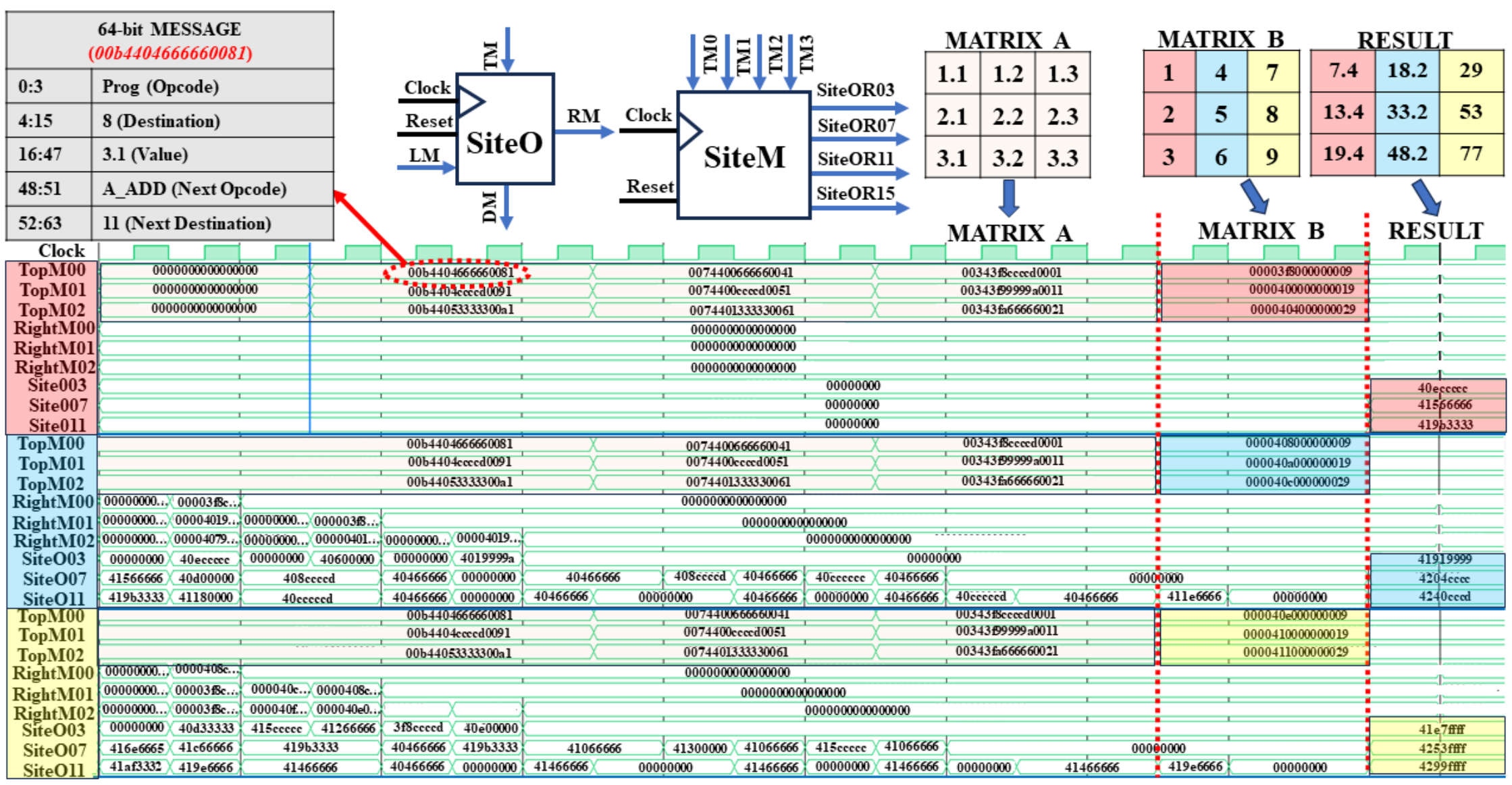

Figure 5: **Simulation-based validation of MAVeC matrix multiplication**. Waveforms illustrate a 3×3 matrix multiplication, showing 64-bit message encoding, SiteO–SiteM interactions, and the temporal sequence of programming, computation, and result generation. Highlighted regions confirm correct operand streaming, in-fabric accumulation, and final output formation.

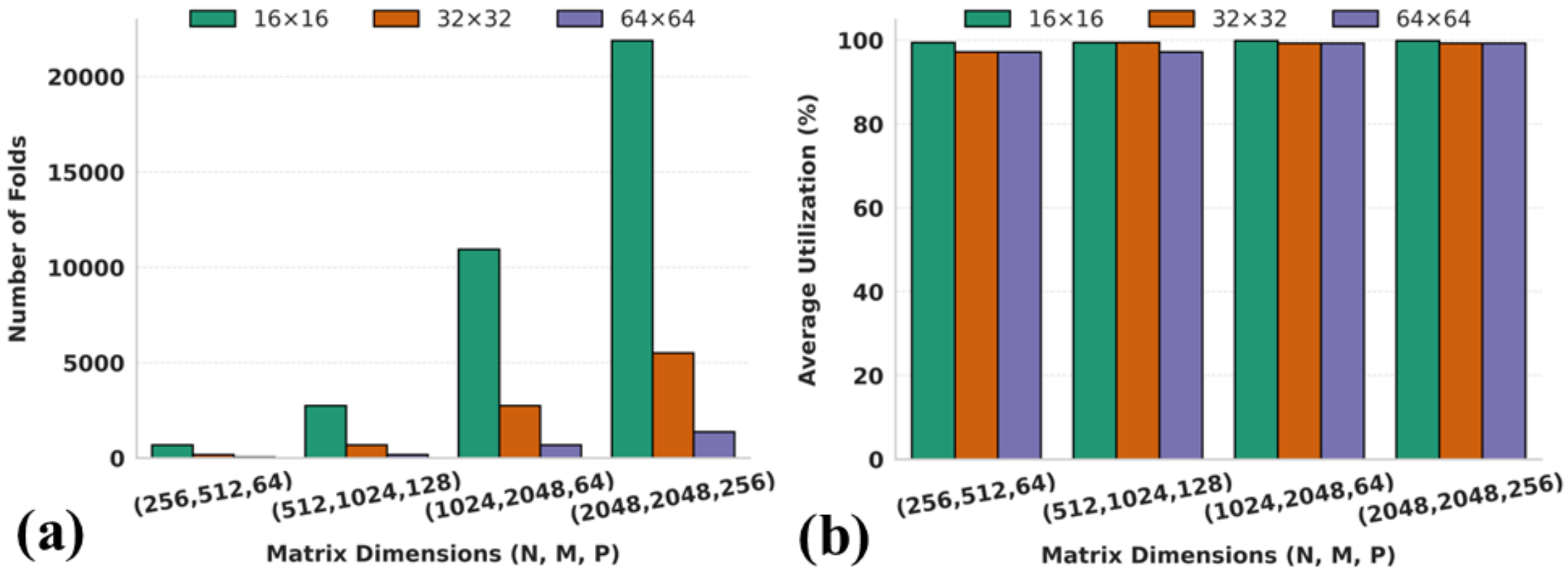

Figure 6: **Average utilization across different hardware and workload configurations. (a)** Number of folds for different configurations, demonstrating reduced folding overhead with increased array size. **(b)** Average utilization across configurations, showing consistently high utilization across array sizes and workloads.

### 6.2 Performance Evaluation

Figure 6 evaluates the effectiveness of MAVeC's mapping strategy using fold statistics and average array utilization. Figure 6(a) plots the total number of generated folds as a function of matrix dimensions (N, M, P) for three SiteO array sizes (16X16, 32X32, and 64X64). As workload size increases, the fold count scales with problem dimensions; however, larger arrays require substantially fewer folds by covering wider spatial regions per fold, thereby reducing folding overhead. Figure 6(b) reports the corresponding average utilization, defined as the fraction of active processing elements across all folds. Despite differences in array size and fold count, utilization remains consistently high ($\geq 97\%$) across all configurations and approaches near-ideal levels for larger matrices, where the impact of partial folds is amortized. Together, these results indicate that MAVeC's interval-aligned folding and message-driven execution translate increased hardware capacity into effective computation rather than idle resources, enabling scalable and efficient utilization across workloads and array sizes.

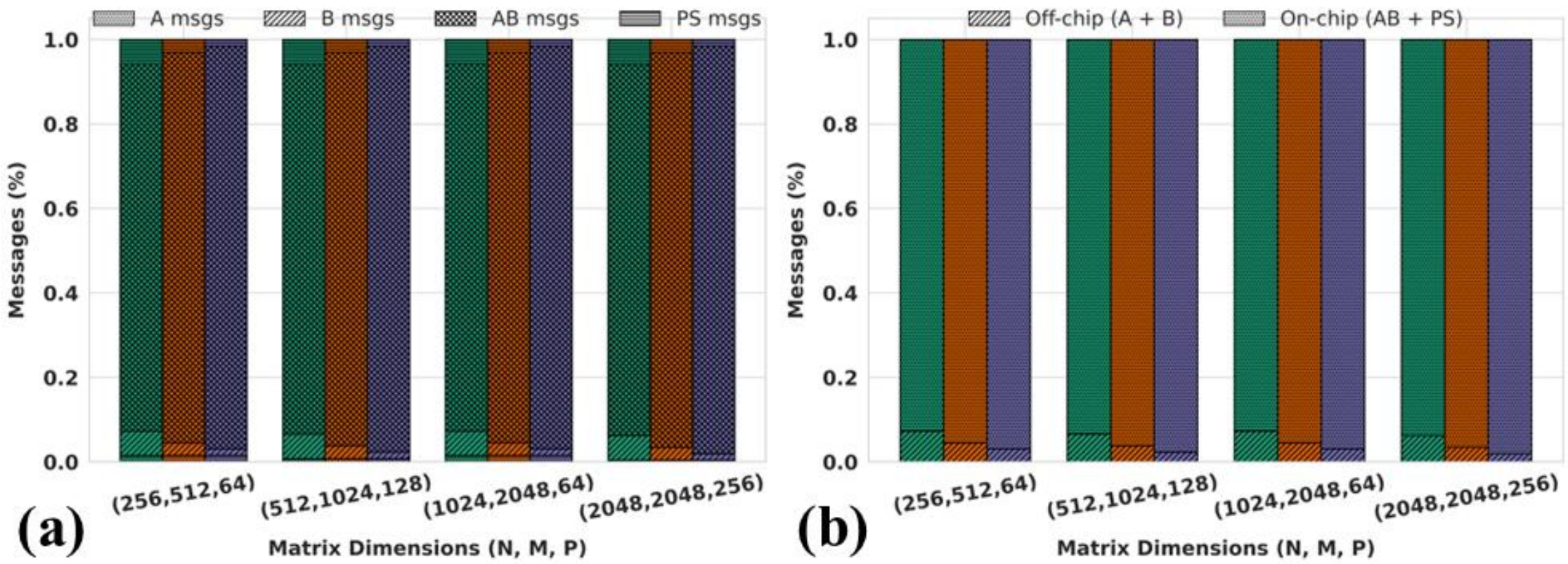

Figure 7: **Message distribution across different hardware and workload configurations. (a)** Normalized breakdown of total messages into input messages (A, B) and intermediate messages (AB, PS). **(b)** Contrasting off-chip traffic (A + B) with on-chip traffic (AB + PS), highlighting communication remains within the MAVeC fabric across workloads.

Figure 7 characterizes MAVeC's communication behavior by analyzing the distribution and locality of messages across matrix dimensions (N, M, P). In Figure 7(a), the x-axis lists workload sizes, while the y-axis shows the normalized fraction of total messages, decomposed into input messages (A, B) and intermediate on-chip messages (AB, PS). AB messages correspond to intermediate multiplication results generated from Matrix-A and Matrix-B operands, whereas PS messages represent partial-sum propagation during accumulation. Across all configurations, intermediate messages dominate ($> 90\%$), indicating that most communication occurs within the fabric after operand injection. Figure 7(b) further groups messages by location, contrasting off-chip traffic (A + B) with on-chip traffic (AB + PS). Off-chip messages account for only a small fraction ($\approx 5-7\%$) of total communication, while most communication is sustained through on-chip message propagation. As matrix sizes grow, this imbalance becomes more pronounced, reflecting improved amortization of input transfers. Overall, these results show that MAVeC effectively confines communication to the on-chip fabric, reducing external bandwidth.

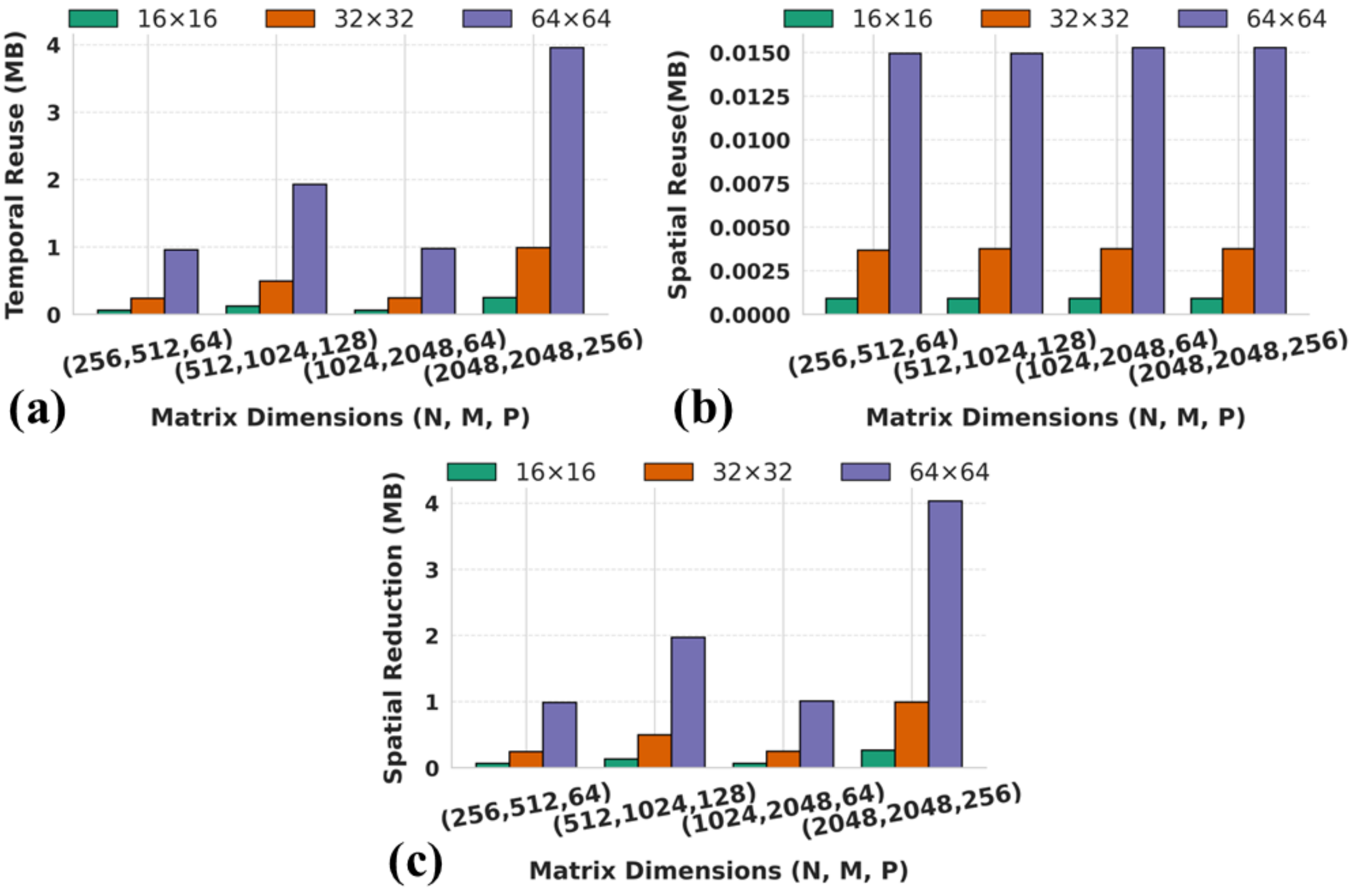

Figure 8: **Reuse and reduction benefits across different hardware and workload configurations. (a)** Average temporal reuse from stationary Matrix-A folds increase with array size. **(b)** Average spatial reuse from multicast of Matrix-B rows, largely independent of workload size and proportional to array height. **(c)** Average spatial reduction from on-chip partial-sum aggregation, yielding higher data-movement savings for larger workloads and wider arrays.

Figure 8 quantifies data-movement benefits from MAVeC's reuse mechanisms across matrix dimensions (N, M, P) and SiteO array sizes (16X16, 32X32, 64X64). In Figure 8(a), the y-axis reports average temporal reuse (MB), showing that reuse increases with both workload size and array width, reaching up to ~4 MB for 64X64 arrays at (2048, 2048, 256) as Matrix-A folds remain stationary across all P Matrix-B folds. Figure 8(b) reports average spatial reuse (MB), which remains nearly constant across workloads but scales with array height, reflecting multicast of Matrix-B folds across $R_P$ processing rows. Figure 8(c) shows average spatial reduction (MB), highlighting substantial savings from on-chip partial-sum aggregation; for large matrices, reduction exceeds 4 MB on 64X64 arrays, indicating effective elimination of redundant partial-sum traffic. Collectively, these results show that MAVeC translates array scaling into proportional reductions in off-chip data movement through coordinated temporal reuse, spatial multicast, and on-fabric reduction.

Figure 9 characterizes the clock-cycle behavior of MAVeC across hardware scales and matrix workloads. Figure 9(a) plots total execution cycles (in million clock cycles) versus matrix dimensions (N, M, P) for 16X16, 32X32, and 64X64 arrays. Execution cycles decrease consistently with increasing array size across all workloads, indicating effective exploitation of spatial parallelism. Figure 9(b) decomposes execution time into data propagation, computation, and partial-sum merge. The results show that data propagation accounts for the largest fraction of runtime across all configurations, ranging from approximately 50% to over 95%, while computation becomes more prominent only for larger problem sizes; partial-sum merge remains a minor overhead (≤3%). Figure 9(c) further breaks down data propagation into its components, revealing that weight propagation dominates data movement (~85-86%), with input (A, B) message propagation contributing substantially smaller shares. Collectively, these results highlight that MAVeC's performance is primarily data-movement-bound rather than compute-bound, emphasizing the architectural importance of efficient on-chip communication and weight-stationary execution for scalable, bandwidth-efficient AI accelerators.

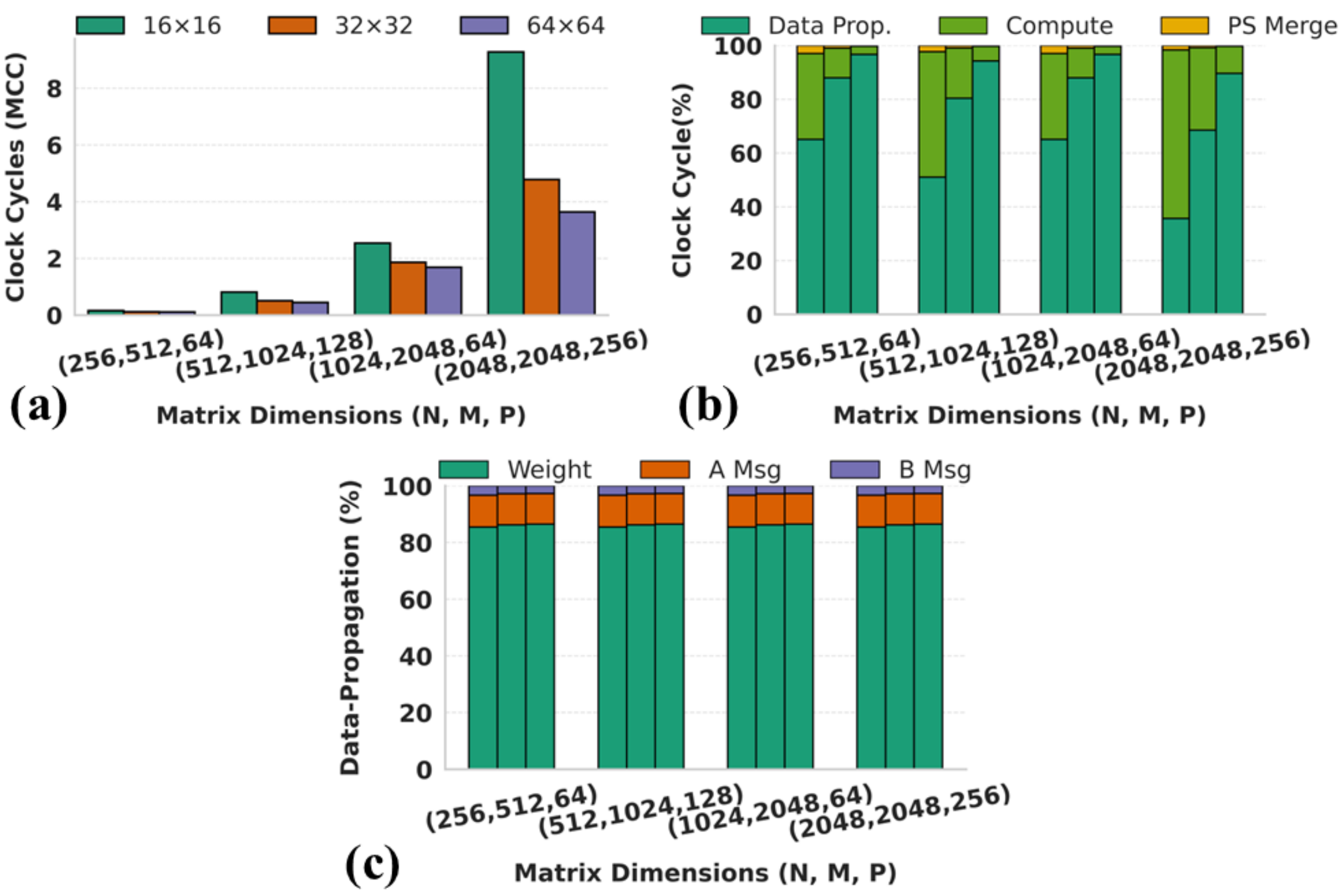

Figure 9: **Clock-cycle characterization across different hardware and workload configurations. (a)** Total execution cycles (in million CCs) decrease with larger arrays due to increased parallelism. **(b)** Clock-cycle breakdown shows that data propagation dominates execution time across all configurations. **(c)** Breakdown of data propagation reveals that weight propagation constitutes the largest share of data movement.

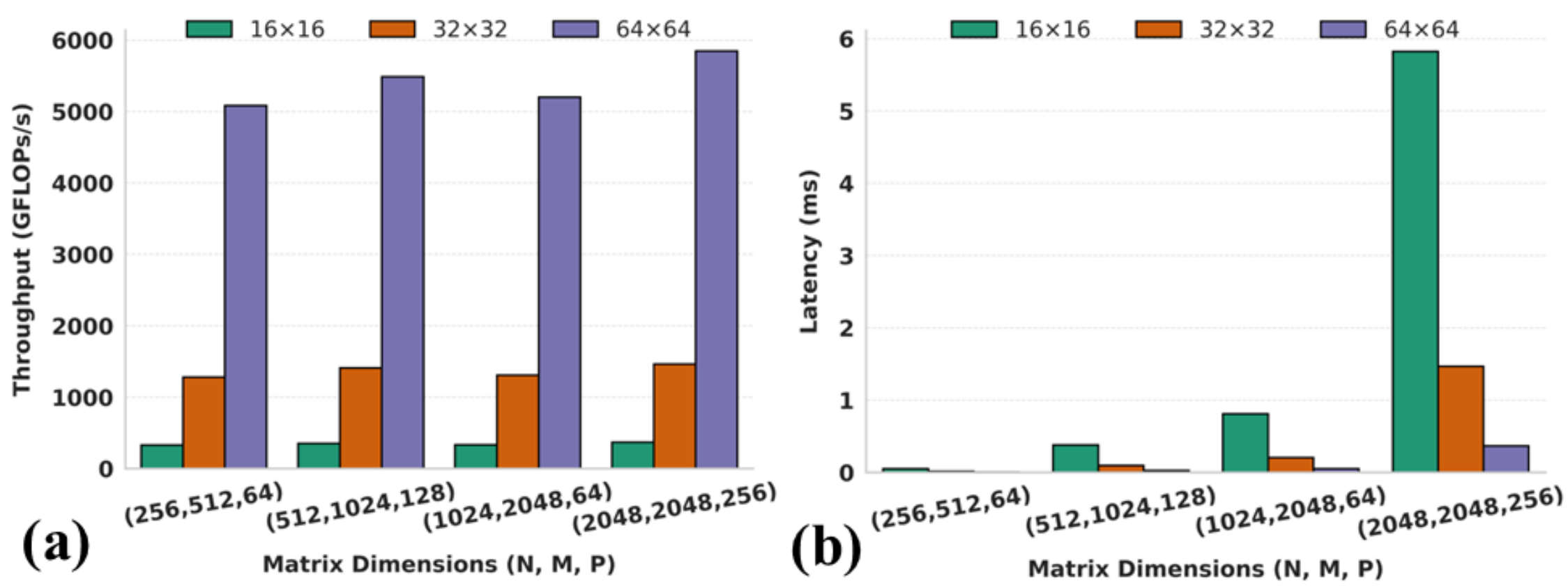

Figure 10: **Throughput and latency across different hardware and workload configurations. (a)** Achieved throughput (GFLOPs/s) reflects near-linear scaling with array size due to increased spatial parallelism. **(b)** End-to-end latency (ms) shows latency reduction with larger arrays as increased parallelism amortizes computation and data-movement overheads.

Figure 10(a) reports throughput (GFLOPs/s) as a function of matrix dimensions (N, M, P) for three SiteO array sizes (16X16, 32X32, and 64X64). The x-axis enumerates workload sizes, while the y-axis shows achieved throughput, with each bar corresponding to a different array configuration. Throughput scales monotonically with array size, increasing from a few hundred GFLOPs/s for 16X16 to over 5 TFLOPs/s for 64X64, demonstrating effective exploitation of spatial parallelism. Figure 10(b) presents the corresponding end-to-end latency in milliseconds. Latency decreases consistently with larger arrays, with the 64X64 configuration reducing latency by more than an order of magnitude compared to 16X16 for large workloads (e.g.,2048X2048X256). Together, these results indicate that MAVeC efficiently converts increased hardware parallelism into higher sustained throughput and lower latency, highlighting the scalability of its message-driven execution model for large matrix workloads.

Figure 11 evaluates MAVeC's energy efficiency across increasing workload sizes and array scales. In Figure 11(a), total energy increases with problem size but decreases with larger arrays for a given workload, indicating that higher spatial parallelism reduces energy per operation by shortening execution time. Figure 11(b) shows that computation dominates energy consumption across all configurations, while data movement-primarily weight propagation-forms the largest non-compute component; message propagation and partial-sum merge remain minor contributors. Figure 11(c) reports average power, which increases with array size due to higher active resources, yet this increase is offset by reduced runtime, leading to lower total energy for larger arrays. Collectively, these results demonstrate that MAVeC effectively trades higher instantaneous power for improved energy efficiency by converting array-level parallelism into reduced execution time and amortized data-movement costs.

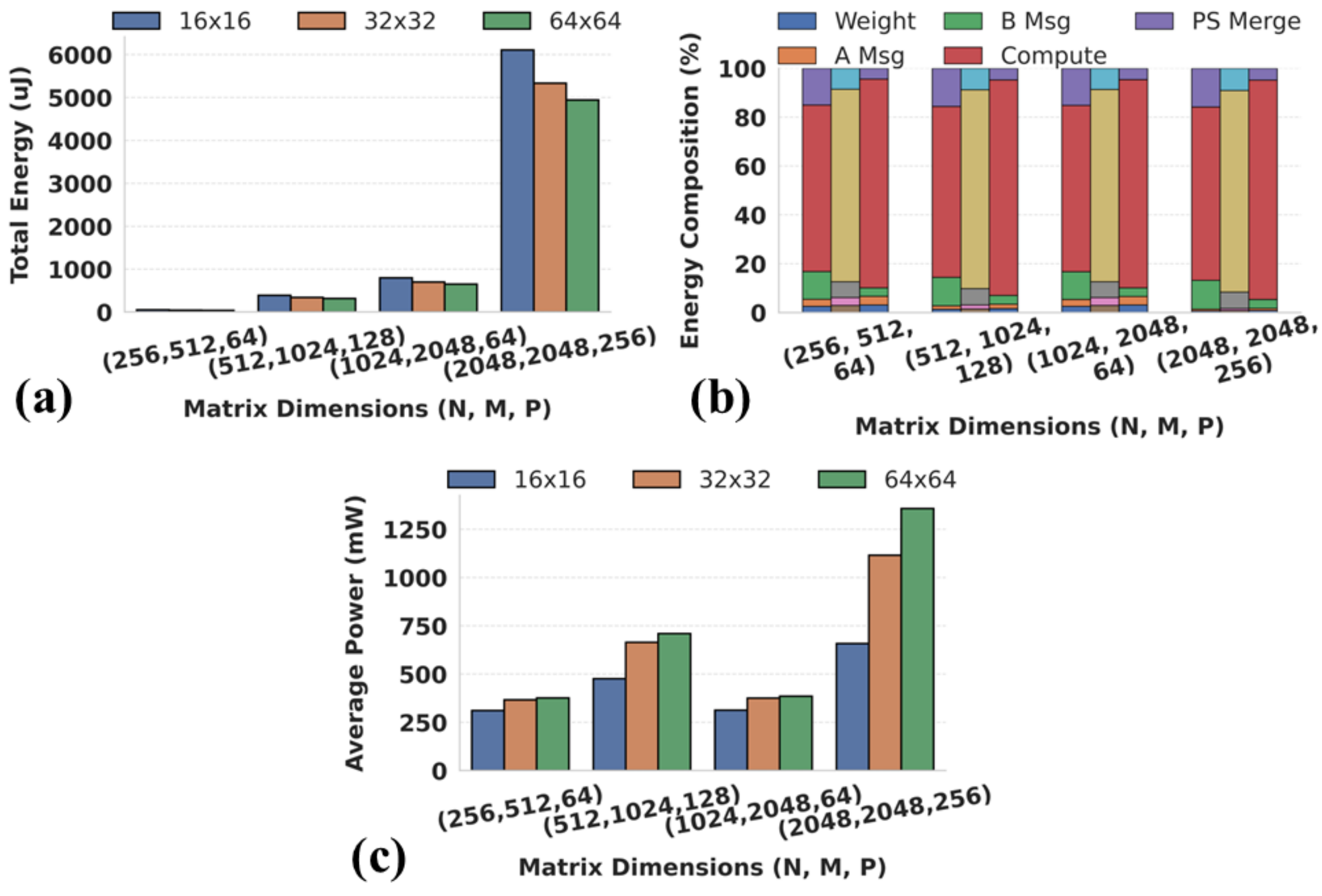

Figure 11: **Energy and power characterization across hardware and workload configurations. (a)** Total energy consumption (μJ) versus matrix dimensions (N, M, P) for 16X16, 32X32, 64X64 SiteO arrays. **(b)** Energy composition breakdown decomposing total energy into weight propagation, message propagation, computation, and partial-sum merge. **(c)** Average power consumption (mW) across configurations, highlighting scaling trends with array size and workload.

Figure 12 presents the layer-wise throughput (GFLOP/s) and average utilization for all convolution layers of VGG-19, evaluated across three MAVeC hardware configurations (16X16, 32×X32, and 64X64). The figure jointly visualizes computational throughput (left axis, bars) and average utilization (right axis, markers), enabling a direct examination of how execution efficiency evolves with both network depth and hardware scale. For the 16×16 configuration, throughput remains relatively modest, ranging from approximately 280–385 GFLOP/s across layers. While most layers achieve near-peak utilization, the first convolution layer (c01) exhibits a noticeable utilization drop (∼75%), which corresponds to a reduced number of folds and partial occupancy caused by small input channel dimensions. This behavior is consistently reflected in the lower throughput observed for this layer, highlighting the sensitivity of smaller arrays to early-layer dimensionality mismatches.

Scaling to 32×32, throughput increases substantially, reaching ∼1.5 TFLOP/s for most layers. Except for the first layer, utilization remains at or near 100%, indicating that the larger array effectively amortizes folding overheads and sustains full spatial occupancy once channel dimensions increase. The tight clustering of throughput values across layers demonstrates stable execution efficiency under sufficient workload parallelism. Again, 64×64 configuration further amplifies throughput, achieving sustained performance between ∼6.0–6.1 TFLOP/s for most layers. Again, utilization remains saturated for nearly all layers, with the only deviation occurring at c01, where utilization drops to approximately 56% due to limited fold availability. Despite this early-layer inefficiency, the overall trend confirms that MAVeC effectively exploits spatial parallelism in deeper layers, where fold counts scale proportionally with array capacity. Overall, Figure 12 shows that throughput increases with array size when the hardware is fully utilized. When utilization drops—most notably in the early convolution layers due to limited fold availability—throughput decreases accordingly. This relationship highlights the importance of dimensional alignment in maintaining high utilization and demonstrates that MAVeC achieves near-ideal performance scaling across most layers of the VGG-19 convolution pipeline.

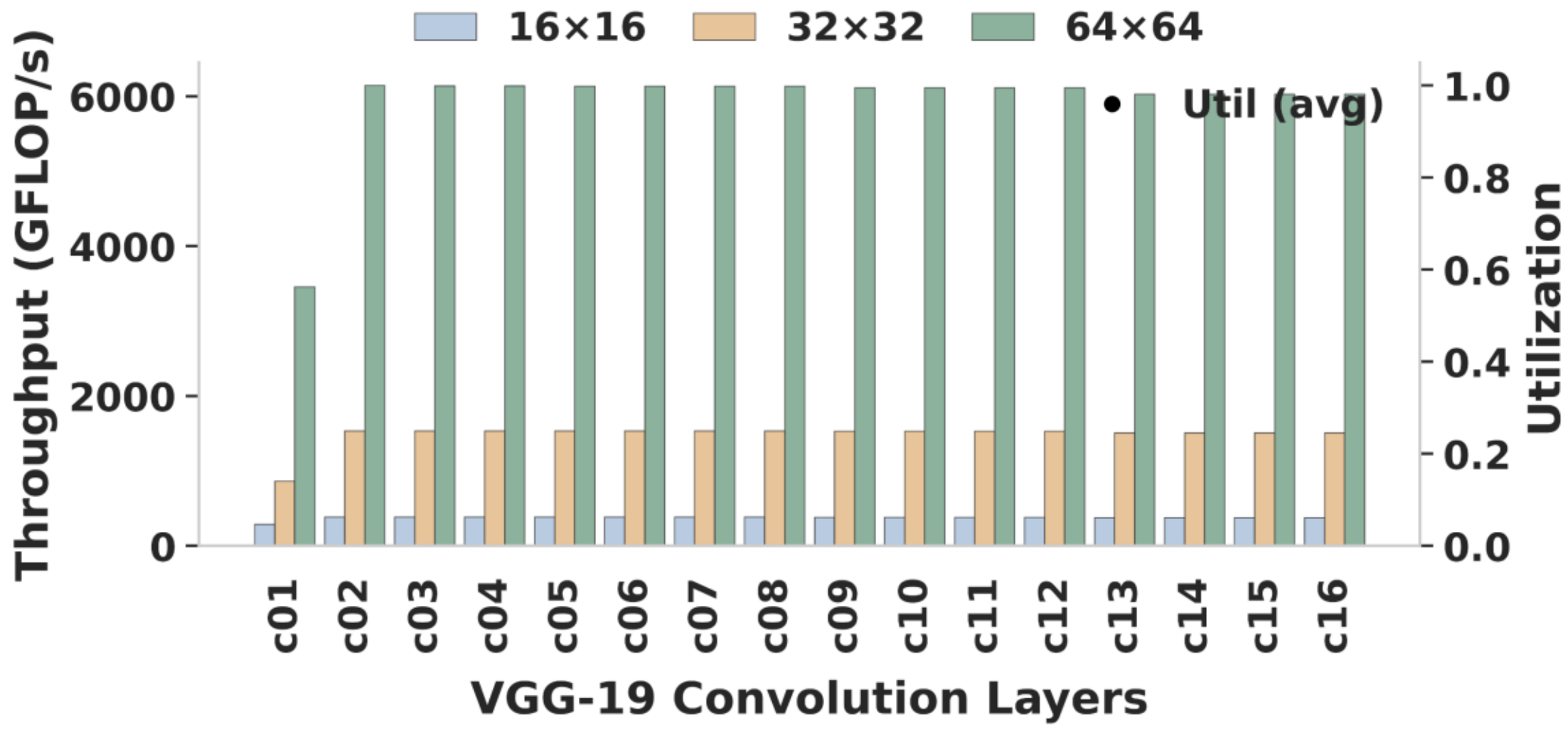

Figure 12: **Layer-wise convolution throughput and average utilization for VGG-19 on MAVeC**. The figure reports per-layer throughput (GFLOP/s) and average utilization across three MAVeC array configurations (16X16, 32X32, and 64X64). Bars denote throughput (left axis), while markers indicate utilization (right axis), highlighting how array scaling and fold availability influence execution efficiency across convolution layers.

### 6.3 Comparison

Table 7 summarizes the analytical characteristics of matrix-matrix multiplication between matrix A (NXM) and matrix B (MXP) across three representative accelerator designs: TPU-style systolic arrays, MEISSA, and MAVeC. The comparison adopts a compute-centric model that assumes sufficient hardware resources to map the full workload without tiling and abstracts away memory hierarchy and bandwidth effects. MEISSA follows a systolic organization similar to TPUs but decouples multipliers from adders while preserving a weight-stationary (WS) dataflow. Under this model, Figure 13(a) evaluates latency scaling by independently varying N, M, and P from 4 to 2048 while holding the remaining dimensions constant at 128. Across all parameter sweeps, MAVeC consistently achieves lower latency than both TPU and MEISSA, with up to a 1.5–2x reduction for large matrix dimensions. This advantage stems from MAVeC's message-driven execution. Instead of propagating non-stationary operands sequentially across rows, MAVeC multicasts operands and performs reduction via decoupled on-fabric aggregation. Thus, accumulation serialization is minimized, and computation overlaps more effectively with reduction, allowing MAVeC to better exploit available compute resources.

Table 7: Summary of matrix multiplication approach between TPU, MEISSA, and MAVeC

| Type | Processing Element | Resource Utilization | Method | Matrix Load | | Latency |
|---|---|---|---|---|---|---|
| | | | | A | B | |
| TPU | MAC | Multiplier: NXP<br>Adders: MXP | Systolic | Store | Left→Right | N + 2M + P -2 |
| MEISSA | Multipliers Adder Tree | Multiplier: MXP<br>Adder: PX(M-1) | Systolic | Store | Left →Right | N + M + P + log (M) - 2 |
| MAVeC | SiteO | {(NXM)+N} X P | Message | Program | Vertical Bus | N + P + 2 |

Figure 13(b) compares total clock cycles for matrix-multiplication workloads executed on 64X64 TPU WS (Weight Stationary dataflow), TPU DiP (Diagonal-Input and Permutated weight stationary dataflow) [1], and MAVeC arrays under different modeling assumptions. TPU-WS and TPU-DiP report compute-centric latency using INT8 precision, assuming preloaded operands and analytically amortized data movement. In contrast, MAVeC evaluates FP32 execution using an end-to-end, system-level model that explicitly captures weight propagation, message routing and multicast, computation, partial-sum merging, and on-fabric data movement. Under this more comprehensive model and higher arithmetic precision, MAVeC reports approximately 1.3-1.6x higher total cycle counts than TPU-based baselines, as operand propagation, synchronization, and reduction costs are explicitly exposed rather than analytically hidden. The observed difference therefore reflects modeling scope and precision rather than reduced compute efficiency.

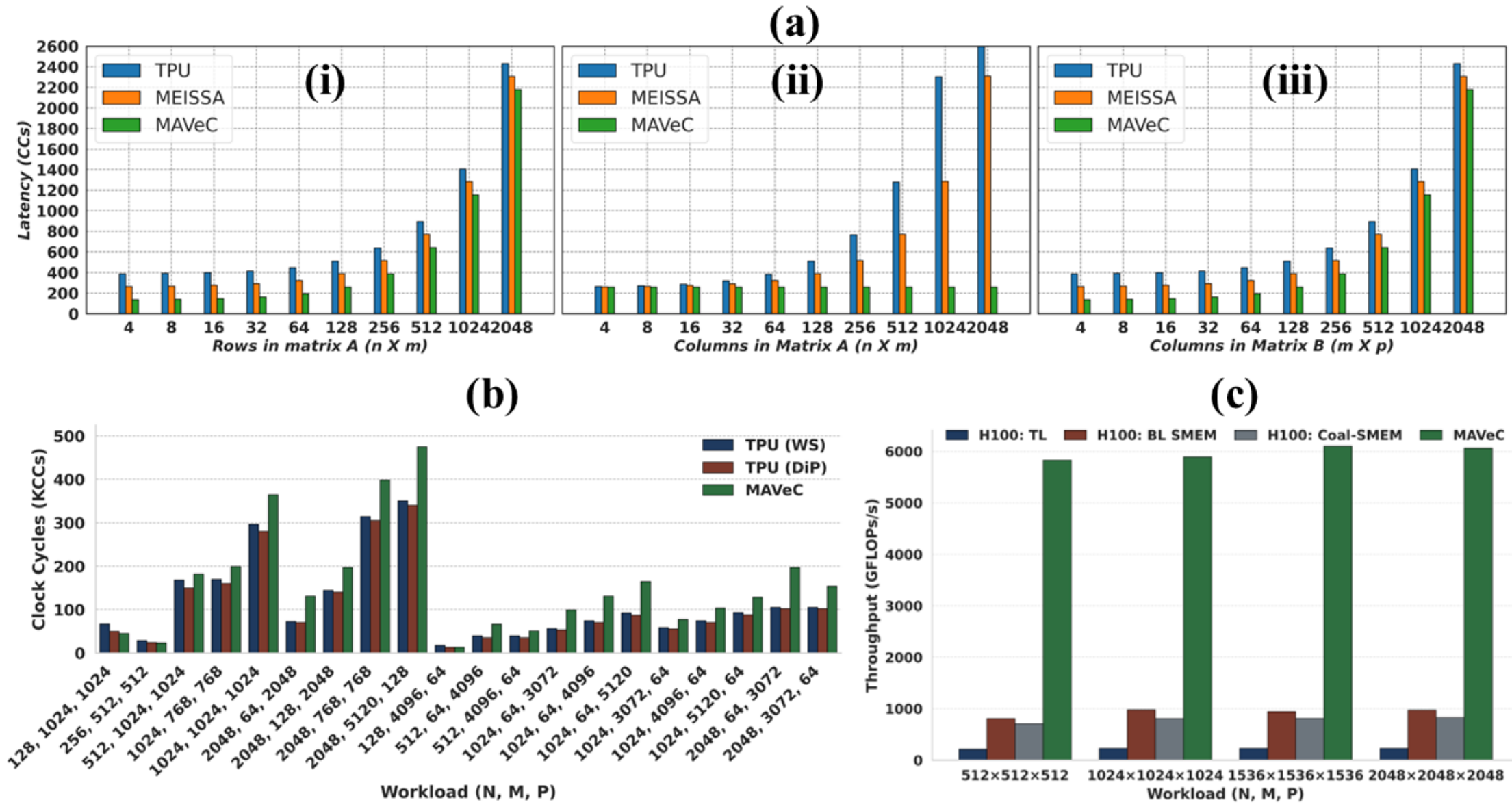

Figure 13: **Comparison of matrix-multiplication performance across architectures. (a)** Latency comparison under independent sweeping of N, M, and P dimensions. **(b)** Clock cycle comparison between systolic and message-driven execution model. **(c)** System-level throughput comparison between MAVeC and NVIDIA H100 GPU.

Figure 13(c) compares system-level FP32 GEMM throughput between NVIDIA H100 GPU kernels[17] and MAVeC. The GPU results correspond to three optimized CUDA implementations: thread-level (TL), block-level shared-memory (BL-SMEM), and coalesced shared-memory (Coal-SMEM) kernels, and include end-to-end effects of global memory access, shared-memory usage, synchronization, and runtime scheduling. In contrast, MAVeC throughput is obtained from a cycle-accurate system-level model that explicitly captures weight propagation, message routing and multicast, computation, and partial-sum merging. Across all evaluated matrix sizes, the GPU saturates at approximately 0.8-1.0 TFLOPs/s for the strongest BL-SMEM kernel, whereas MAVeC sustains 5.8-6.1 TFLOPs/s. This yields a consistent 6.0-7.2x throughput advantage for MAVeC. Moreover, MAVeC's throughput remains nearly constant as problem size increases, indicating effective balancing of computation and data movement under its message-driven execution model, while GPU performance remains bounded by memory hierarchy behavior and synchronization overheads despite tiling and shared-memory optimization.

## 7 CONCLUSION

The growing dominance of data movement, synchronization, and control overheads in modern AI workloads exposes fundamental limitations in accelerator architectures that rely on centralized orchestration or rigid dataflows. This work addressed these limitations through MAVeC, a messaging-based adaptive vector computing architecture that embeds execution, routing, and coordination directly into the hardware fabric. By exploiting the deterministic parallelism of AI workloads, MAVeC eliminates centralized scheduling and accumulation serialization, allowing computation, data movement, and reduction to progress autonomously via self-propagating messages. Our simulation-based results lend credibility; over 90% of within chip communications are autonomous. Comparative evaluation results show gain in performance due to this decoupling from centralized execution; in the best-case scenario greater than 6x throughput benefits were observed for various matrix workloads with MAVeC over state-of-the-art GPUs. The performance gains are achieved without compromising scalability or energy efficiency, as shorter execution times amortize data-movement costs even at higher instantaneous power, establishing message-driven execution as a scalable and effective foundation for future AI accelerators.

## ACKNOWLEDGMENTS

The authors received no financial support for the research, authorship, and publication of this article.